\documentclass[copyright, creativecommons]{eptcs}
\providecommand{\event}{F-IDE 2016} 
\usepackage{breakurl}               

\usepackage[utf8]{inputenc}

\usepackage{amssymb}
\usepackage{xcolor}

\usepackage{pbox}
\usepackage{relsize}
\usepackage{subfigure}
\usepackage{wrapfig}
\usepackage{pdflscape}

\usepackage[inline]{enumitem}

\usepackage{graphicx}
  \DeclareGraphicsExtensions{.png,.pdf,.jpg,.jpeg}
  \graphicspath{{./}{./pictures/}{./pictures/cisco-icons/}{./pictures/oxygen-icons/}{./figures/}}
\usepackage{tikz}
  \usetikzlibrary{positioning}
  \usetikzlibrary{shapes.geometric,shapes.symbols}
  \usetikzlibrary{automata}
\usepackage{msc5}

\usepackage{letltxmacro} 

\usepackage{comment}
\includecomment{otherToolsDescription}
\excludecomment{FLangFullSpec}
\excludecomment{FLangFullTest}

\usepackage[style=bubble,tight,condAfterRULE,inlinedSyntax]{k} 
  \renewcommand{\kconfig}[1]{\ensuremath{#1}} 
  \LetLtxMacro{\oldKrule}{\krule}
  \renewcommand{\krule}[6][]{%
    \hspace*{\stretch{1}}%
    \nolinebreak[4]%
    \fbox{\pbox{\linewidth}{\oldKrule[#1]{#2}{#3}{#4}{#5}{#6}}}%
    \nolinebreak[4]%
    \hspace*{\stretch{1}}%
    \linebreak[0]%
  }
  \LetLtxMacro{\oldKall}{\kall}
  \renewcommand{\kall}[3][white]{%
    \hspace*{\stretch{1}}%
    \nolinebreak[4]%
    \oldKall[#1]{#2}{#3}%
    \nolinebreak[4]%
    \hspace*{\stretch{1}}%
    \linebreak[0]%
  }%
  \LetLtxMacro{\oldKprefix}{\kprefix}
  \renewcommand{\kprefix}[3][white]{%
    \hspace*{\stretch{1}}%
    \nolinebreak[4]%
    \oldKprefix[#1]{#2}{#3}%
    \nolinebreak[4]%
    \hspace*{\stretch{1}}%
    \linebreak[0]%
  }%
  \LetLtxMacro{\oldKsuffix}{\ksuffix}
  \renewcommand{\ksuffix}[3][white]{%
    \hspace*{\stretch{1}}%
    \nolinebreak[4]%
    \oldKsuffix[#1]{#2}{#3}%
    \nolinebreak[4]%
    \hspace*{\stretch{1}}%
    \linebreak[0]%
  }%
  \LetLtxMacro{\oldKmiddle}{\kmiddle}
  \renewcommand{\kmiddle}[3][white]{%
    \hspace*{\stretch{1}}%
    \nolinebreak[4]%
    \oldKmiddle[#1]{#2}{#3}%
    \nolinebreak[4]%
    \hspace*{\stretch{1}}%
    \linebreak[0]%
  }%
  \LetLtxMacro{\oldReduce}{\reduce}
  \renewcommand{\reduce}[2]{%
    \hspace*{\stretch{1}}%
    \nolinebreak[4]%
    \oldReduce{#1}{#2}%
    \nolinebreak[4]%
    \hspace*{\stretch{1}}%
    \linebreak[0]%
  }%

  \newenvironment{kConfDef}[1][0.9]{
    \begin{kdefinition}%
      \relscale{#1}%
  }{%
    \end{kdefinition}%
  }
  \newenvironment{kSemDef}[1][0.9]{
    \begin{kdefinition}%
      \relscale{#1}%
  }{%
    \end{kdefinition}%
  }

\usepackage{listings}

\definecolor{mygreen}{rgb}{0,0.6,0}
\definecolor{mygray}{rgb}{0.5,0.5,0.5}
\definecolor{mymauve}{rgb}{0.58,0,0.82}

\lstdefinelanguage{K}{
  morekeywords={require,imports,module,endmodule,syntax,configuration,rule},
  morekeywords={[2]strict,left,right,structural,bracket},
  morekeywords={[3]List,Id,Int,Bool,String},
  morekeywords={[4]when,requires},
  morekeywords={[5]notBool,in,keys},
  sensitive=true,
  morestring=[b]",
  morecomment=[l]{//},
}
\lstdefinestyle{K}{
  escapeinside={¤}{¤},
  breaklines=true,
  language=K,
  showstringspaces=false,
  basicstyle=\smaller[0]\ttfamily,
  keywordstyle={\bfseries\color{green!40!black}},
  keywordstyle={[2]\mdseries\color{blue!60!black}},
  keywordstyle={[3]\mdseries\color{green!30!black}},
  keywordstyle={[4]\mdseries\color{red!60!black}},
  keywordstyle={[5]\mdseries\color{green!30!black}},
  identifierstyle={},
  stringstyle=\color{orange},
  commentstyle=\itshape\color{purple!40!black},
}
\lstdefinelanguage{FLang}{
  morekeywords={PROLOGUE,AUTOMATA,INIT,FILTER},
  morekeywords={[2]cond,iter,newInterrupt},
  morekeywords={[3]set,newAutomaton,step,alarm,nop},
  morekeywords={[4]accept,drop,send},
  sensitive=true,
  morestring=[b]",
  morecomment=[l]{//},
}
\lstdefinestyle{FLang}{
  tabsize=8,
  breaklines=true,
  xleftmargin=\parindent,
  language=FLang,
  showstringspaces=false,
  basicstyle=\smaller[1]\ttfamily,
  keywordstyle={\bfseries\color{green!40!black}},
  keywordstyle={[2]\mdseries\color{green!60!black}},
  keywordstyle={[3]\mdseries\color{blue!60!black}},
  keywordstyle={[4]\mdseries\color{red!60!black}},
  identifierstyle={},
  stringstyle=\color{orange},
  commentstyle=\itshape\color{purple!40!black},
  escapeinside=\£,
}

\title{
  Industrial Experience Report on the 
  Formal Specification of a Packet Filtering Language 
  Using the K Framework
}
\def\titlerunning{Formal Specification of a Packet Filtering Language Using the K Framework}

\author{
  Gurvan \textsc{Le Guernic}
  \institute{
    DGA Maîtrise de l'Information \\
    35998 Rennes Cedex 9, France
  }
  \and
  Benoit \textsc{Combemale}
  \quad\qquad
  José A. \textsc{Galindo}
  \institute{
    INRIA RENNES – BRETAGNE ATLANTIQUE\\
    Campus universitaire de Beaulieu \\
    35042 Rennes Cedex, France
  }
}
\def\authorrunning{G. Le Guernic, B. Combemale \& J.A. Galindo}

\newcommand{\FLang}[0]{GPFL}

\begin{document}

\maketitle

\begin{abstract}
  Many project-specific languages, including in particular filtering languages, are defined using non-formal specifications written in natural languages. This leads to ambiguities and errors in the specification of those languages.
  This paper reports on an industrial experiment on using a tool-supported language specification framework ({\K}) for the formal specification of the syntax and semantics of a filtering language having a complexity similar to those of real-life projects.
  This experimentation aims at estimating, in a specific industrial setting, the difficulty and benefits of formally specifying a packet filtering language using a tool-supported formal approach.
\end{abstract}

\section{Introduction} \label{sec:introduction}

 Packet filtering (accepting, rejecting, modifying or generating packets, i.e. strings of bits, belonging to a sequence) is a recurring problematic in the domain of information systems security. Such filters can serve, among other uses, to reduce the attack surface by limiting the capacities of a communication link to the legitimate needs of the system it belongs to. This type of filtering can be applied to network links (which is the most common use), product interfaces, or even on the communication buses of a product. If the filtering policy needs to be adapted during the deployment or operational phases of the system or product, it is often required to design a specific language $\mathcal{L}$ (syntax and semantics) to express new filtering policies during the lifetime of the system or product. This language is the basis of the filters that are applied to the system or product. Hence, it plays an important role in the security of this system or product. It is therefore important to have strong guarantees regarding the expressivity, precision, and correctness of the language $\mathcal{L}$ (meaning that everything that need to be expressed can, and that everything that can be expressed has the most obvious semantics). Those guarantees can be partly provided by a formal design (and development) process.

 Among diverse duties, the DGA (Direction Générale de l’Armement, a french procurement agency) is involved in the supervision of the design and development of filtering components or products. Those filters come in varying shapes and roles. Some of them are network apparatuses filtering standard Internet protocol packets (such as firewalls); while others are small parts of integrated circuits filtering specific proprietary packets transiting on computer buses. Their common definition is: ``a tool sitting on a communication channel, analyzing the sequence of packets (strings of bits with a beginning and an end) transiting on that channel, and potentially dropping, modifying or adding packets in that sequence''. Whenever the filtering algorithm applied is fixed for the lifetime of the component or product, this algorithm is often ``hard coded'' into the component or product with the potential addition of a configuration file allowing to slightly alter the behavior of the filter. However, sometimes the filtering algorithm to apply may depend on the deployment context, and may have to evolve during the lifetime of the component or product to adapt to new uses or attackers. In this case, it is often necessary to be able to easily write new filtering algorithms for the specific product and context. Those algorithms are then often described using a Domain Specific Language (DSL) that is designed for the expression of a specific type of filters for a specific product. The definition of the syntax and semantics of this DSL is an important task. This DSL is the link between the filtering objectives and the process that is really applied on the packet sequences.
 Often, language specifications (when there is one) are provided using natural language. In the majority of cases, this leads to ambiguities or errors in the specification which propagate to implementations and final user code. This is for example the case for common languages such as C/C++ or Java{\texttrademark}~\cite{Sun:2004:JSR133}.
 \begin{quote}
   ``Unfortunately, the current specification has been found to be hard to understand and has subtle, often unintended, implications. Certain synchronization idioms sometimes recommended in books and articles are invalid according to the existing specification. Subtle, unintended implications of the existing specification prohibit common compiler optimizations done by many existing Java virtual machine implementations. [...] Several important issues, [...] simply aren’t discussed in the existing specification.'' \\
  \hspace*{\stretch{1}} JSR-133 expert group \cite{Sun:2004:JSR133}
 \end{quote}
 Some of those ambiguities, as the memory model of multi-threaded Java{\texttrademark} programs \cite{Sun:2004:JSR133}, required a formal specification in order to be solved.

 This paper is an industrial experience report on the use of a tool-supported language specification framework (the {\K} framework) for the formal specification of the syntax and semantics of a filtering language having a complexity similar to those of real-life projects.
 The tool used to formally specify the DSL is introduced in Sect.~\ref{sec:intro-K}.
 For confidentiality reasons, in order to be allowed by the DGA to communicate on this experimentation, the language specified for this experiment is not linked to any particular product or component. It is a generic packet filtering language that tries to cover the majority of features required by packet filtering languages. This language is introduced in Sect.~\ref{sec:flang-context} while its formal specification is described in Sect.~\ref{sec:flang-formal-specification}. This language is tested in Sect.~\ref{sec:testing} by implementing and simulating a filtering policy enforcing a sequential interaction for a made-up protocol similar to DHCP. Before concluding in Sect.~\ref{sec:conclusion}, this paper discusses the results of the experimentation in Sect.~\ref{sec:retex}.

\section{Introduction to the \K{} Framework}
\label{sec:intro-K}

 Surprisingly, even if it is a niche for tools, there exists quite a number of tools specifically dedicated to the formal \emph{specification} of languages (our focus in this work is on specifying rather than implementing DSLs). Those tools include among others: PLT Redex \cite{felleisen:2009:SEwPR,klein:2012:ryr}, Ott \cite{sewell:2010:Ott}, Lem \cite{mulligan:2014:Lem}, Maude MSOS Tool \cite{chalub:2007:MMT}, and the \K{} framework \cite{rosu:2010:K-overview,serbanuta:2014:K-primer}. All those tools focus on the (clear formal) specification of languages rather than their (efficient) implementation, which is more the focus of tools and languages such as Rascal \cite{klint:2009:RASCAL,basten:2015:mlir,klint:2011:RLSRunner} or its ancestor The Meta-Environment \cite{klint:2009:t2gmt,storm:2008:Meta-Env}, Kermeta \cite{jezequel:2011:mdlewK,jezequel:2013:mm}, and others.
\begin{otherToolsDescription}
   PLT Redex is based on reduction relations. PLT Redex is an extension (internal DSL) of the Racket programming language~\cite{flatt:2010:racket-reference}.
   Ott and Lem are more oriented towards theorem provers. Ott and Lem allow to generate formal definitions of the language specified for Coq, HOL, and Isabelle. In addition, Lem can generate executable OCaml code. Ott is more programming language syntax oriented, while Lem is a more general purpose semantics specification tool. Ott and Lem can be used together in some contexts.
   The Maude MSOS Tool, whose development has stopped in 2011, is based on an encoding of modular structural operational semantics (MSOS) rules into Maude.
   Similarly to the Maude MSOS Tool, the \K{} framework is based on rewriting and was also originally implemented on top of Maude. 
\end{otherToolsDescription}

 The goal set for the experiment reported in this paper is to
%
 estimate
%
 the difficulty and benefits for an average engineer (i.e. an engineer with education and experience in computer science but no specific knowledge in formal language semantics) to use an ``appropriate'' tool for the formal specification of a packet filtering language.
 The ``appropriate'' tool needs to: be easy to use; be able to produce (or take as input) ``human readable'' language specifications; provide some level of correctness guarantees for the language specified; and be executable (simulatable) in order to test (evaluate) the language specified. The \K{} framework seems to meet those requirements and has been chosen to be the ``appropriate'' tool after a short review of available tools. As there has been no in depth comparison of the different tools available, there is no claim in this paper that the \K{} framework is better than the other tools, even in our specific setting.

 This section introduces the \K{} framework \cite{rosu:2014:K-overview} by relying on the example of a language allowing to compute additions over numbers using Peano's encoding \cite{peano:1889:tpoa}. The \K{} source code of this language specification is provided below. 
%
\begin{lstlisting}[style=K,escapechar=¤,basicstyle={\smaller[0]\ttfamily}]
module PEANO-SYNTAX
  syntax Nb ::= "Zero" | "Succ" Nb   ¤\label{K:peano:syntax:Nb}¤
  syntax Exp ::= Nb | Id | Exp "+" Exp     [strict,left]
  syntax Stmt ::= Id ":=" Exp ";"          [strict(2)]
  syntax Prg ::= Stmt | Stmt Prg
endmodule

module PEANO imports PEANO-SYNTAX
  syntax KResult ::= Nb                        ¤\label{K:peano:syntax:KResult}¤

  configuration
    <env color="green"> .Map </env>
    <k color="cyan"> $PGM:K </k>

  rule N:Nb + Zero => N                        ¤\label{K:peano:rule:addZero}¤
  rule N1:Nb + Succ N2:Nb => ( Succ N1 ) + N2  ¤\label{K:peano:rule:addSucc}¤

  rule                                         ¤\label{K:peano:rule:var}¤
    <env> ... Var:Id |-> Val:Nb ... </env>
    <k> ( Var:Id => Val:Nb ) ... </k>

  rule
    <env> Rho:Map (.Map => Var |-> Val ) </env>
    <k> Var:Id := Val:Nb ; => . ... </k>
    when notBool (Var in keys(Rho))

  rule
    <env> ... Var |-> ( _ => Val ) ... </env>
    <k> Var:Id := Val:Nb ; => . ... </k>

  rule S:Stmt P:Prg => S ~> P [structural]     ¤\label{K:peano:rule:seq}¤
endmodule
\end{lstlisting}
%
 A \K{} definition is divided into three parts: the \emph{syntax} definition, the \emph{configuration} definition, and the \emph{semantics} (rewriting rules) definition.
 The definition of the language \emph{syntax} is given in a module whose name is suffixed with ``-SYNTAX''. It uses a BNF-like notation \cite{backus:1959:BNF,knuth:1964:BNFvsBNF}. Every non-terminal is introduced by a \texttt{syntax} rule. For example, the definition of the notation for numbers (\texttt{Nb}) in this language, provided on line~\ref{K:peano:syntax:Nb}
, is equivalent to the definition given by the regular expression ``(\texttt{Succ})\textsuperscript{*} \texttt{Zero}''.
%

 \begin{wrapfigure}{R}{14em}
   \vspace*{-1ex}
   \centering

   \input{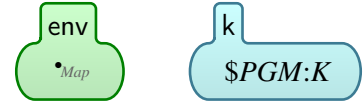}

   \caption{Peano's \K{} configuration}
   \label{fig:peano:k-configuration}

   \vspace*{-1ex}
 \end{wrapfigure}

 The \emph{configuration} definition part is introduced by the keyword \texttt{configuration} and defines a set of (potentially nested) cells described in an XML-like syntax. This configuration describes the ``abstract machine'' used for defining the semantics of the language. The initial state (or configuration) of the abstract machine is the one described in this configuration part. The parsed program (using the syntax definition of the previous part) is put in the cell containing the \texttt{\$PGM} variable (of type \texttt{K}). For the \texttt{Peano} language, the \texttt{env} cell is used to store variable values in a map initially empty (\texttt{.Map} is the empty map). From this definition, the \K{} framework can produce a graphical representation of the configuration, provided in Fig.~\ref{fig:peano:k-configuration}

 The \emph{semantics} definition part is composed of a set of rewriting rules, each one of them introduced by the keyword \texttt{rule}. In the \K{} source file, rules are roughly denoted as ``$CCF$ \texttt{=>} $NCF$'' where $CCF$ and $NCF$ are configuration fragments. The meaning of ``$CCF$ \texttt{=>} $NCF$'' can be summarized as: if $CCF$ is a fragment of the current abstract machine state (or configuration) then the rule may apply and the fragment matching $CCF$ in the current configuration would then be replaced by the new configuration fragment $NCF$. In order to increase the expressivity of rules, $CCF$ may contain free variables that are reused in expressions in $NCF$. If a specific valuation of the free variables $V$ in $CCF$ allows a fragment of the current configuration to match $CCF$, then this fragment may be replaced by $NCF$ where the variables $V$ are replaced by their matching valuation.

 The rules for addition over numbers (\texttt{Nb} and not \texttt{Exp}), on lines~\ref{K:peano:rule:addZero} and~\ref{K:peano:rule:addSucc}
, follows closely this representation. For those rules, $CCF$ is a program fragment that can be matched in any cell of the configuration. For those two rules,  the \K{} framework can then produce the following graphical representations:\\[1ex]
\hspace*{\stretch{1}}
\mbox{
  \krule{
    \reduce
      {{\variable[Nb]{N}{user}}\terminal{+}{{}\terminal{Zero}}}
      {\variable[Nb]{N}{}}
  }{}{}{}{}
}
\hspace*{\stretch{1}}
\mbox{
  \krule{
    \reduce
      {{\variable[Nb]{N1}{user}}\terminal{+}{{}\terminal{Succ}{\variable[Nb]{N2}{user}}}}
      {{\left({{}\terminal{Succ}{\variable[Nb]{N1}{}}}\right)}\terminal{+}{\variable[Nb]{N2}{}}}
  }{}{}{}{}
}
\hspace*{\stretch{1}}

 For other rules, the configuration fragment matching is more complex and involves precise configuration cells that are explicitly identified. In order to compress the representation, $CCF$ and $NCF$ are not stated separately anymore. The common parts are stated only once, and the parts differing are again denoted ``$CCF_i$ \texttt{=>} $NCF_i$'', where $CCF_i$ is a sub-fragment in $CCF$ and $NCF_i$ is the corresponding sub-fragment in $NCF$. Cells that have no impact on a rule $R$ and are not impacted by $R$ do not appear explicitly in the rule. Cells heads and tails (potentially empty) that are not modified by a rule can be denoted ``\texttt{...}'', instead of using a free variable that would not be reused.

 \begin{wrapfigure}{R}{17em}
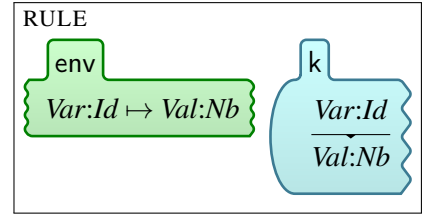

   \vspace*{-1ex}
   \centering
   \krule{
     \kmiddle[green]{env}{
       {\variable[Id]{Var}{user}}\mapsto{\variable[Nb]{Val}{user}}
     }
     \mathrel{}
     \kprefix[cyan]{k}{
       \reduce
       {\variable[Id]{Var}{user}}
       {\variable[Nb]{Val}{user}}
     }
   }{}{}{}{}
   \caption{Peano's \K{} rule for variables}
   \label{fig:peano:k-rule-getVar}
   \vspace*{-1ex}
 \end{wrapfigure}

 For example, the rule which starts on line~\ref{K:peano:rule:var}
 is the rule used to evaluate variables. The current configuration needs to contain a mapping from a variable \texttt{Var} to a value \texttt{Val} (``\texttt{X |-> V}'' denotes a mapping from \texttt{X} to \texttt{V}) somewhere in the map contained in the \texttt{env} cell. It also needs to contain the variable \texttt{Var} at the beginning of cell \texttt{k}. This rule has the effect of replacing the instance of \texttt{Var} at the beginning of cell \texttt{k} by the value \texttt{Val}. For this rule, the \K{} framework generates the graphical representation given in Fig.~\ref{fig:peano:k-rule-getVar}.

 The last rule on line~\ref{K:peano:rule:seq} involves other internal aspects of the \K{} framework. It roughly states that, in order to evaluate a statement \texttt{S} followed by the rest \texttt{P} of the program, \texttt{S} must first be evaluated to a \texttt{KResult} (defined on line ~\ref{K:peano:syntax:KResult}) and then \texttt{P} is evaluated.


\section{\FLang{} Context}
\label{sec:flang-context}

 The language specified in the experiment reported in this paper, named \FLang{}, is a generic packet filtering language. For confidentiality reasons, \FLang{} is not a language actually used in any specific real product. \FLang{} has been made-up in order to be able to communicate on the experimentation on tool supported formal specification of filtering languages reported in this paper. However, \FLang{} covers the majority of features needed in packet filtering languages dealt with by the DGA. \FLang{} can be seen as the ``mother'' of the majority of packet filtering languages.

 \FLang{} aims at expressing a wide variety of filters. Those filters can be placed at the level of network, interfaces, or even communication buses between electronic components. They can be applied on standard protocols such as IP, TCP, UDP, \dots{} or on proprietary protocols, which are more common for component communication protocols. However, all those filters are assumed to be placed on a communication link. Messages (packets) that get through the filter can only get through in two ways, either ``going in'' or ``going out''; there is no switching taking place in \FLang{} filters. Those different use cases are illustrated in Fig.~\ref{fig:filter-use-cases}.

\newcommand{\filterIcon}{guard}
\def\iconSpaceAtOutSide{2em}%
\def\iconSpaceAtInSide{1.4em}%
\newcommand{\filterDirection}[1]{%
  \ifthenelse{\equal{#1}{left2right}}{%
    \def\filterOutSide{west}%
    \def\filterInSide{east}%
    \edef\spaceBeforFilterIcon{\iconSpaceAtInSide}%
    \edef\spaceAfterFilterIcon{\iconSpaceAtOutSide}%
  }{%
    \def\filterOutSide{east}%
    \def\filterInSide{west}%
    \edef\spaceBeforFilterIcon{\iconSpaceAtOutSide}%
    \edef\spaceAfterFilterIcon{\iconSpaceAtInSide}%
  }%
}
\filterDirection{left2right}%
\newcommand{\drawFilter}[2][Filter]{%
  \begin{scope} \smaller
    \node[rectangle, draw] (#1) at #2 {
      \hspace*{\spaceBeforFilterIcon} \includegraphics[width=4em]{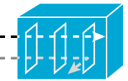} \hspace*{\spaceAfterFilterIcon}
    };
    \node[rectangle, draw, anchor=\filterOutSide] (#1InI)  at ($(#1.north \filterOutSide)!0.33!(#1.south \filterOutSide)$) {in};
    \node[rectangle, draw, anchor=\filterInSide] (#1OutI) at ($(#1.south \filterInSide)!0.33!(#1.north \filterInSide)$) {out};
    \node[anchor=south] (#1Label) at (#1.north) {\FLang{} filter};
  \end{scope}
}
\newcommand{\drawFilterConnections}[3][Filter]{%
  \begin{scope}[-latex, thick, font={\smaller[3]}]
    \coordinate (inShift) at ($(#1InI.\filterOutSide) - (#1.\filterOutSide)$);
    \draw[red] ([shift={(inShift)}] #2.\filterInSide) -- ([shift={(inShift)}] #1.\filterOutSide);
    \draw[red, loosely dashed] ([shift={(inShift)}] #1.\filterInSide) -- ([shift={(inShift)}] #3.\filterOutSide);
    
    \coordinate (outShift) at ($(#1OutI.\filterInSide) - (#1.\filterInSide)$);
    \draw[blue] ([shift={(outShift)}] #3.\filterOutSide) -- ([shift={(outShift)}] #1.\filterInSide);
    \draw[blue, densely dashed] ([shift={(outShift)}] #1.\filterOutSide) -- ([shift={(outShift)}] #2.\filterInSide);
  \end{scope}
}

 \begin{figure}[!htb]
   \centering
   
   \newcommand{\monitoringDeviceNtwIcon}{128x128_apps_esd}
   \newcommand{\desktopNtwIcon}{256x256_mimetypes_application-x-smb-server}
   \newcommand{\InternetNtwIcon}{256x256_mimetypes_application-x-smb-workgroup}
   \newcommand{\databaseNtwIcon}{256x256_places_network-server-database}
   \newcommand{\serverNtwIcon}{256x256_places_network-server}
   \subfigure[Network filtering]{
     \begin{tikzpicture}
       \drawFilter{(0,0)}%
    
       \node[anchor=east] (outNetwork) at ([xshift=-2em] Filter.west) {%
         \includegraphics[height=3em]{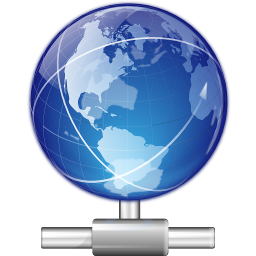}%
         \includegraphics[height=3em]{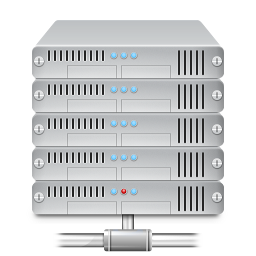}%
         \includegraphics[height=3em]{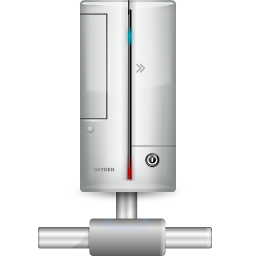}%
       };
       \node[anchor=west] (inNetwork)  at ([xshift=2em] Filter.east) {%
         \includegraphics[height=3em]{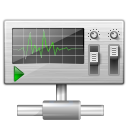}%
         \includegraphics[height=3em]{\serverNtwIcon}%
         \includegraphics[height=3em]{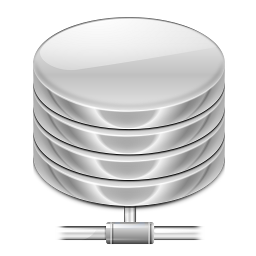}%
       };

       \drawFilterConnections{outNetwork}{inNetwork}
     \end{tikzpicture}
   }

   \newcommand{\desktopItfIcon}{256x256_devices_computer}
   \newcommand{\keyboarbItfIcon}{256x256_devices_input-keyboard}
   \newcommand{\USBKeytfIcon}{256x256_devices_drive-removable-media-usb-pendrive}
   \subfigure[Interface filtering]{

     \begin{tikzpicture}
       \node (Computer) {\includegraphics[height=3em]{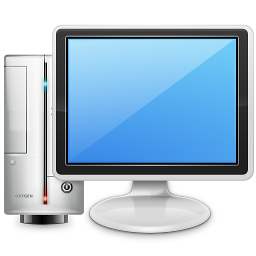}};

       \filterDirection{left2right}%
       \drawFilter[KeyboardFilter]{([xshift=-6em] Computer.west)}%
       \filterDirection{right2left}%
       \drawFilter[USBKeyFilter]{([xshift=6em] Computer.east)}%

       \node[anchor=east] (Keyboard) at ([xshift=-2em] KeyboardFilter.west) {%
         \includegraphics[height=3em]{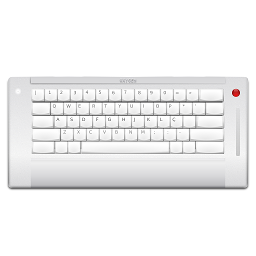}%
       };
       \node[anchor=west] (USBKey) at ([xshift=2em] USBKeyFilter.east) {%
         \includegraphics[height=3em]{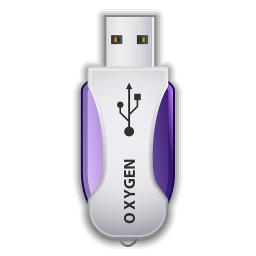}%
       };

       \filterDirection{left2right}%
       \drawFilterConnections[KeyboardFilter]{Keyboard}{Computer}
       \filterDirection{right2left}%
       \drawFilterConnections[USBKeyFilter]{USBKey}{Computer}
     \end{tikzpicture}
   }

   \newcommand{\secProcCmpIcon}{128x128_devices_cpu.png}
   \newcommand{\memChipCmpIcon}{256x256_devices_media-flash}
   \newcommand{\memCardCmpIcon}{256x256_devices_media-flash-smart-media}
   \subfigure[Bus filtering]{
     \begin{tikzpicture}

       \node[anchor=east] (MemChip) {%
         \includegraphics[height=3em]{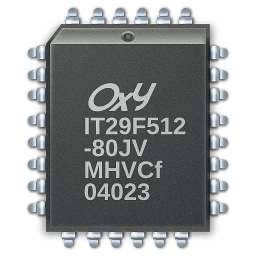}%
       };

       \drawFilter[FirstFilter]{([xshift=-6em] MemChip.west)}%

       \drawFilter[SecondFilter]{([xshift=6em] MemChip.east)}%

       \node[anchor=east] (MemCard) at ([xshift=-2em] FirstFilter.west) {%
         \includegraphics[height=3em]{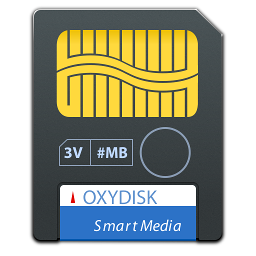}%
       };

       \node[anchor=west] (CPU)  at ([xshift=2em] SecondFilter.east) {%
         \includegraphics[height=3em]{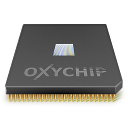}%
       };

       \drawFilterConnections[FirstFilter]{MemCard}{MemChip}
       \drawFilterConnections[SecondFilter]{MemChip}{CPU}
     \end{tikzpicture}
   }

   \caption{Use cases for \FLang{}-based filters}
   \label{fig:filter-use-cases}
 \end{figure}

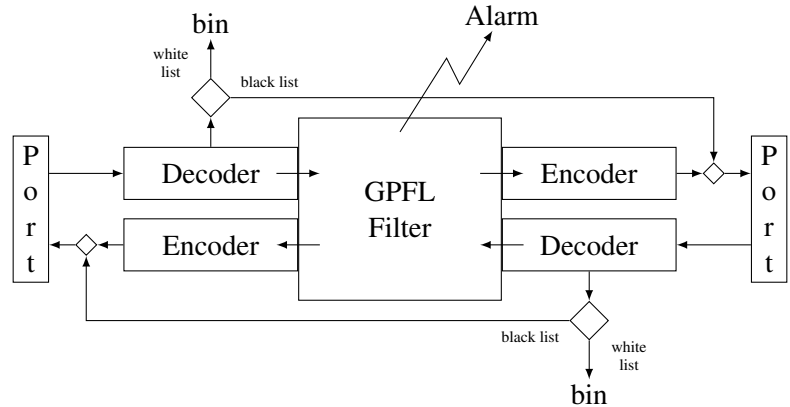
\begin{figure}[!htb]
   \centering
   
   \tikzset{
     port/.style={draw, rectangle, align=center},
     filter/.style={draw, rectangle, align=center},
     errorHandler/.style={draw, rectangle},
     coder/.style={draw, rectangle, minimum width=6em, minimum height=4ex},
     encoder/.style={coder},
     decoder/.style={coder},
     choice/.style={draw, diamond},
     bin/.style={inner sep=2pt, outer sep=0pt},
     merge/.style={draw, diamond, inner sep=2pt},
     alarm/.style={inner sep=2pt, outer sep=0pt},
   }

   \hspace*{\stretch{1}}
   \begin{tikzpicture}
     \node[filter, minimum height=14ex, minimum width=7em] (filter) {\FLang{} \\ Filter};
     \node[alarm, above=3em of filter.north east, anchor=south] (alarm) {Alarm};
     \node[decoder, anchor=south east] (inDecoder) at ($(filter.north west)!.45!(filter.south west)$) {Decoder};
     \node[encoder, anchor=west] (inEncoder) at ($(filter.north east)!(inDecoder.center)!(filter.south east)$) {Encoder};
     \node[decoder, anchor=north west] (outDecoder) at ($(filter.south east)!.45!(filter.north east)$) {Decoder};
     \node[encoder, anchor=east] (outEncoder) at ($(filter.south west)!(outDecoder.center)!(filter.north west)$) {Encoder};
     \path (inDecoder) |- coordinate (tmpCoord) (filter.north);
     \node[choice, above=0ex of tmpCoord] (choiceIn) {};
     \path (outDecoder) |- coordinate (tmpCoord) (filter.south);
     \node[choice, below=0ex of tmpCoord] (choiceOut) {};
     \node[bin, above=3ex of choiceIn] (binIn) {bin};
     \node[bin, below=3ex of choiceOut] (binOut) {bin};
     \node[merge, right=2ex of inEncoder](mergeIn){};
     \node[merge, left=2ex of outEncoder](mergeOut){};
     \path (filter) -| coordinate (tmpCoord) (mergeOut.west);
     \node[port, left=2ex of tmpCoord] (portL) {P\\o\\r\\t};
     \path (filter) -| coordinate (tmpCoord) (mergeIn.east);
     \node[port, right=2ex of tmpCoord] (portR) {P\\o\\r\\t};

     \begin{scope}[-latex]
       \path (portL.east) |- coordinate (tmpCoord) (inDecoder); \draw (tmpCoord) -- (inDecoder);
       \draw ([xshift=-.75em] inDecoder.east) -- ([xshift=.75em] inDecoder.east);
       \draw (inDecoder) -- (choiceIn);
       \draw (choiceIn) -- (binIn) node[pos=0,anchor=south east,xshift=-.75em,inner sep=0pt,font={\smaller[3]},align=center]{white \\ list};
       \draw (choiceIn) -| (mergeIn) node[at start,above,anchor=south west,font={\smaller[3]}]{black list};
       \draw ([xshift=-.75em] inEncoder.west) -- ([xshift=.75em] inEncoder.west);
       \draw (inEncoder) -- (mergeIn);
       \path (mergeIn) -| coordinate (tmpCoord) (portR.west); \draw (mergeIn) -- (tmpCoord);

       \path (portR.west) |- coordinate (tmpCoord) (outDecoder); \draw (tmpCoord) -- (outDecoder);
       \draw ([xshift=.75em] outDecoder.west) -- ([xshift=-.75em] outDecoder.west);
       \draw (outDecoder) -- (choiceOut);
       \draw (choiceOut) -- (binOut) node[pos=0,anchor=north west,xshift=.75em,inner sep=0pt,font={\smaller[3]},align=center]{white \\ list};
       \draw (choiceOut) -| (mergeOut) node[at start,below,anchor=north east,font={\smaller[3]}]{black list};
       \draw ([xshift=.75em] outEncoder.east) -- ([xshift=-.75em] outEncoder.east);
       \draw (outEncoder) -- (mergeOut);
       \path (mergeOut) -| coordinate (tmpCoord) (portL.east); \draw (mergeOut) -- (tmpCoord);

       \path ([yshift=-.5em] filter.north) -- coordinate[pos=.6] (tmpCoord) (alarm);
       \draw ([yshift=-.5em] filter.north) -- ([shift={(-.5ex,1ex)}] tmpCoord) -- ([shift={(.5ex,-1ex)}] tmpCoord) -- (alarm);
     \end{scope}

   \end{tikzpicture}
   \hspace*{\stretch{1}}

   \caption{Architecture of \FLang{}-based filters}
   \label{fig:filter-architecture}
 \end{figure}

 \FLang{} focuses on the internal logic of the filter. Decoding and encoding of packets is assumed to be handled outside of \FLang{} programs (filters), potentially using technologies such as ASN.1 \cite{ISO/IEC:8824-1:2015,dubuisson:2000:ASN1}. For \FLang{} programs, a packet is a record (a set of valued fields). A \FLang{} program (dynamically) inputs a sequence of records and outputs a sequence of records. Figure~\ref{fig:filter-architecture} describes the architecture of \FLang{}-based filters. An incoming packet (on either side) is first parsed (decoded) before being handed over to the \FLang{} program. If the packet can not be parsed, depending on the type of filter (white list or black list), the packet is either dropped or passed to the other side without going through the \FLang{} program. Any packet (record) output by the \FLang{} program (on either side) is encoded before being sent out. In addition, the \FLang{} program can generate alarms due to packets not complying with the encoded filtering policy.

 The \FLang{} language must allow to: drop, modify or accept the current packet being filtered; generate new packets; and generate alarms%
. \FLang{} must allow to base the decision to take any of those actions on information pieces concerning the current packet being filtered and previously filtered packets. Those information pieces must include: some timing information, current or previous packets directions through the filter (``in'' or ``out''), and characteristics of current or previous packets including field values and computed properties such as, for example, a packet ``type'' or total length. The computation of those properties and decoding of packet fields is outside of the scope of \FLang{}; it is left to the decoders.

 In order to gradually build a decision, \FLang{} must allow to interact with variables (reading, writing, and computing expressions) and automata (triggering a transition in an automaton and querying its current state). The intent for automata is to be used to track the current step of sessions of complex protocols. \FLang{} must allow to combine filtering statements using: sequential control statements (executing two statements in sequence); conditional control statements (executing a statement only if a condition is true); iterating control statements (repeatedly executing a statement for a fixed number of repetitions). There is no requirement for a loop (or while) statement whose exit condition is controlled by an expression recomputed after every iteration. For the experiment reported in this paper (on formal specification of a filtering language), the iterating statement is considered sufficient for the intended use of \FLang{} and close enough to a loop statement from a semantics point of view, while exhibiting interesting properties for future analyses (for example, any \FLang{} program terminates).

\section{\FLang{}'s Specification}
\label{sec:flang-formal-specification}

 Due to lack of space, {\FLang}'s specification and testing is only summarized in this paper. However, a full specification of {\FLang} and a testing section can be found in the companion technical report~\cite{le-guernic:2016:erfspflukf}. 

 \paragraph{Syntax.} \label{sec:flang-formal-syntax}
 To the exception of expressions and expression fragments, {\FLang}'s syntax is formally defined by the \K{} source fragment provided below.
\begin{lstlisting}[style=K,firstline=18,lastline=40,firstnumber=18,basicstyle={\smaller[1]\ttfamily}]
  syntax Cmd ::= "nop" | "accept" | "drop" | "send(" Port "," Fields ")"
              |  "alarm(" Exp ")"                                [strict(1)]
              |  "set(" Id "," Exp ")"                           [strict(2)]
              |  "newAutomaton(" String "," AutomatonId ")"
              |  "step(" AutomatonId "," Exp "," Stmt ")"        [strict(2)]
  syntax Stmt ::= Cmd
               |  "cond(" Exp "," Stmt ")"                       [strict(1)]
               |  "iter(" Exp "," Stmt ")"                       [strict(1)]
               |  "newInterrupt(" Int "," Bool "," Stmt ")"
               |  Stmt Stmt                                      [right]
               |  "{" Stmt "}"                                   [bracket]

  syntax AutomataDef ::= "AUTOMATA" String AutomataDefTail
  syntax AutomataDefTail ::= "init" "=" AStateId ATransitions | ATransitions
  syntax ATransitions ::= List{ATransition,""}
  syntax ATransition ::= AStateId "-" AEvtId "->" AStateId
  syntax AStateId ::= String
  syntax AEvtId ::= String
  syntax InitSeq ::= "INIT" Stmt
  syntax PrologElt ::= AutomataDef | InitSeq
  syntax Prologues ::= PrologElt | PrologElt Prologues

  syntax Program ::= "PROLOGUE" Prologues "FILTER" Stmt
\end{lstlisting}

 A \FLang{} program is composed of a prologue, executed only once in order to initialize the execution environment, and a filter statement, executed once for every incoming packet.
 A prologue is composed of automaton kind definitions and initialization sequences.
 An automaton kind definition specifies an identifier $K$, an initial state for automata of kind $K$ and a set of transitions for automata of kind $K$. A transition definition is composed of: two automaton states $F$ and $T$, and an automaton event that triggers the transition from $F$ to $T$.
    
 A \FLang{} statement is composed of \FLang{} commands or statements combined sequentially. Some statements can be guarded by an expression and executed only if that expression evaluates to true (\texttt{cond}). Some statements (\texttt{iter}), associated with an expression $e$, are exectued $v$ times, where $v$ is the value of $e$ before the first iteration.
 Finally, \texttt{newInterrupt} statements register a statement to be executed in the future, potentially periodically.
    
 \FLang{} commands are the basic units having an effect on the execution environment.
 The \texttt{nop} command has no effect and serves mainly as a place holder.
 The \texttt{accept}, resp. \texttt{drop}, command states to accept, resp. drop, the current packet and stop the filtering process for this packet.
 The \texttt{send} command sends a packet on one of the ports.
 The \texttt{alarm} command generates a message on the alarm channel.
 The \texttt{set} command sets the value of a variable.
 The \texttt{newAutomaton} command initializes an automaton of the provided kind, and assigns this newly created automaton to the provided identifier.
 The \texttt{step} command tries to trigger an automaton transition by sending an event $e$ to an automaton $a$. If there is no transition from the current state of $a$ triggered by the event $e$, then the associated statement is executed.

 \paragraph{Semantics}
 \colorlet{stmtColor}{gray}
 \colorlet{pktColor}{blue}
 \colorlet{timeColor}{teal!75}
 \colorlet{automataColor}{green}
 \colorlet{envColor}{cyan}
 \colorlet{interruptsColor}{orange}
 \colorlet{streamsColor}{olive}
 \colorlet{alarmColor}{red}
 The full formal specification of {\FLang}'s semantics can be found in the companion technical report~\cite{le-guernic:2016:erfspflukf}. {\FLang}'s semantics rules are defined on the configuration presented graphically in Fig.~\ref{fig:gpfpl:k-configuration}.
 \begin{figure}[!htb]
   \centering 
      
   \resizebox{\linewidth}{!}{
     \begin{kConfDef}
       \input{./K-src/GPFPL/GPFPL-CONFIG.k.tex}
     \end{kConfDef}
   }

   \caption{\K{} configuration of \FLang{}}
   \label{fig:gpfpl:k-configuration}
 \end{figure}
 The \texttt{prg} cell contains the \FLang{} program.
 After initialization of the program, automaton kind definitions are stored in the \texttt{automatonKindDefs} cell and the \texttt{filter} cell contains the filter (\FLang{} statement) that is to be executed for every packet.
 The \texttt{interrupts} cell contains a set of interrupt definitions (\texttt{interrupt*}). An interrupt is a triplet composed of: the time when the interrupt is to be triggered, the code (statement) to be executed, and a ``Time'' value equal to the interruption period for a periodic interruption (or nothing for a non-periodic interruption).
 In addition, the \texttt{interrupts} cell contains an ordered list of the next ``times'' when an interrupt is to be executed.
 The \texttt{clock} cell registers the current ``time''.
 The configuration also contains a \texttt{k} cell that holds the {\FLang} statement under execution. Each time a new packet is input, the content of the \texttt{k} cell is replaced by the content of the \texttt{filter} cell, and the newly arrived packet is stored in the \texttt{input} cell with its arrival time and port.

 Packets are input from the \texttt{streams} cell which contains:
 the packet input stream divided into the next packet to arrive (\texttt{inHead}) and the rest of the stream (\texttt{inTail});
 the packet output stream;
 and the alarm output stream.
 In the input stream, resp. output stream, packets arriving, resp. leaving, on both ports are mixed together, but contains information on the port of entry, resp. exit.
 Some choices made to represent those streams are not an intrinsic part of {\FLang}'s formal specification. The division of the input stream into a head and a tail is such a choice. Those choices are made in order to be able to execute the specification. It is then required to implement, in the {\K} framework, a mechanism to retrieve and parse strings describing packet sequences sent to the filter. In order to help distinguish between the formal specification of \FLang{} and the mechanisms put in place to execute it, whenever possible, implementation choices, such as the format of strings describing packets, are defined in another file which is loaded in the main specification file with the \texttt{require} instruction.

 Finally, the \texttt{env} cell is the main dynamic part of the execution environment. It corresponds to a ``record'' of maps that associate: automaton kind and current state to automaton identifiers (\texttt{automata} cell); and values to variables.

\begin{FLangFullSpec}    
    \section{Formal Specification of \FLang{}}
    \label{sec:flang-formal-specification}
    
    \subsection{Syntax}
    \label{sec:flang-formal-syntax}
    
     The syntax of \FLang{} is formally defined by the \K{} source fragment provided in Fig.~\ref{fig:gpfpl:k-src-file:syntax}.
     \begin{figure}[!htbp]
\begin{lstlisting}[style=K,firstnumber=9]
  syntax ExpVal ::= Int | Bool | String | AEvtId | Port

  syntax BuiltInId ::= "_inPort" 
  syntax VarId ::= Id
  syntax FieldId ::= "$" Id
  syntax AutomatonId ::= "#" Id
  syntax ExpId ::= BuiltInId | VarId | FieldId | AutomatonId

  syntax UnaryOp ::= "--" | "!"
  syntax BinaryOp ::= "+" | "-" | "*" | "/" | "&" | "|"
                   | "==" | "<" | ">" | "<=" | ">="

  syntax Exp ::= ExpVal | ExpId
              |  UnaryOp Exp           [strict(2)]          ¤\label{K:GPFPL:syntax:unaryOp}¤
              |  Exp BinaryOp Exp      [strict(1,3), left]  ¤\label{K:GPFPL:syntax:binaryOp}¤
              |  "(" Exp ")"           [bracket]

  syntax Cmd ::= "nop" | "accept" | "drop" | "send(" Port "," Fields ")"
              |  "alarm(" Exp ")"                                [strict(1)]
              |  "set(" Id "," Exp ")"                           [strict(2)]
              |  "newAutomaton(" String "," AutomatonId ")"
              |  "step(" AutomatonId "," Exp "," Stmt ")"        [strict(2)]
  syntax Stmt ::= Cmd
               |  "cond(" Exp "," Stmt ")"                       [strict(1)]
               |  "iter(" Exp "," Stmt ")"                       [strict(1)]
               |  "newInterrupt(" Int "," Bool "," Stmt ")"
               |  Stmt Stmt                                      [right]
               |  "{" Stmt "}"                                   [bracket]

  syntax AutomataDef ::= "AUTOMATA" String AutomataDefTail
  syntax AutomataDefTail ::= "init" "=" AStateId ATransitions | ATransitions
  syntax ATransitions ::= List{ATransition,""}
  syntax ATransition ::= AStateId "-" AEvtId "->" AStateId
  syntax AStateId ::= String
  syntax AEvtId ::= String
  syntax InitSeq ::= "INIT" Stmt
  syntax PrologElt ::= AutomataDef | InitSeq
  syntax Prologues ::= PrologElt | PrologElt Prologues

  syntax Program ::= "PROLOGUE" Prologues "FILTER" Stmt
\end{lstlisting}
       \caption{\K{} source file of \FLang{} syntax}
       \label{fig:gpfpl:k-src-file:syntax}
     \end{figure}
     A \FLang{} program is composed of a prologue, executed only once in order to initialize the execution environment, and a filter statement, executed once for every incoming packet.
    
     A prologue is composed of automaton kind definitions and initialization sequences.
     An automaton kind definition specifies an identifier $K$, an initial state for automata of kind $K$ and a set of transitions for automata of kind $K$. A transition definition is composed of: two automaton states $F$ and $T$, and an automaton event that triggers the transition from $F$ to $T$.
    
     A \FLang{} statement is composed of \FLang{} commands or statements combined sequentially. Some statements can be guarded by an expression and executed only if that expression evaluates to true (\texttt{cond}). Some statements, associated with an expression $e$, can be executed multiple times (\texttt{iter}), as much times as the expression $e$ evaluates to before the first iteration.
     Finally, the \texttt{newInterrupt} statement registers a statement to be executed in the future, potentially periodically.
    
     \FLang{} commands are the basic units having an effect on the execution environment.
     The \texttt{nop} command has no effect and serves mainly as a place holder.
     The \texttt{accept}, resp. \texttt{drop}, command states to accept, resp. drop, the current packet and stop the filtering process for this packet.
     The \texttt{send} command sends a packet on one of the ports.
     The \texttt{alarm} command generates a message on the alarm channel.
     The \texttt{set} command sets the value of a variable.
     The \texttt{newAutomaton} command initializes an automaton of the provided kind and assigns the provided identifier to interact with this newly created automaton.
     The \texttt{step} command tries to trigger an automaton transition by sending an event $e$ to an automaton $a$. If there is no transition from the current state of $a$ triggered by the event $e$, then the associated statement is executed.
    
     Expressions in \FLang{} are quite standard. Primitive values include integers, booleans, strings, automata events and ports. The only ``somewhat'' uncommon aspect of \FLang{} is that automaton identifiers in expressions are evaluated to the current state of the associated automaton.
    
    
    
    \subsection{Configuration}
    
    \colorlet{stmtColor}{gray}
    \colorlet{pktColor}{blue}
    \colorlet{timeColor}{teal!75}
    \colorlet{automataColor}{green}
    \colorlet{envColor}{cyan}
    \colorlet{interruptsColor}{orange}
    \colorlet{streamsColor}{olive}
    \colorlet{alarmColor}{red}
    
     The configuration used to execute \FLang{} programs is presented graphically in Fig.~\ref{fig:gpfpl:k-configuration}.
    %
    %
      \begin{figure}[!htb]
        \centering 
          
          \begin{kConfDef}
            \input{./K-src/GPFPL/GPFPL-CONFIG.k.tex}
          \end{kConfDef}
    
        \caption{\K{} configuration of \FLang{}}
        \label{fig:gpfpl:k-configuration}
      \end{figure}
    %
    %
     A configuration contains a set of
     automaton kind definitions (\texttt{automatonDef}), with the same information as defined in Sect.~\ref{sec:flang-formal-syntax}.
     The \texttt{prg} cell contains the \FLang{} program. After initialization of the program, the \texttt{filter} cell contains the filter (\FLang{} statement) that is to be executed for every packet.
     The \texttt{env} cell is the main dynamic part of the execution environment. It corresponds to a ``record'' of maps that associate:
     automaton kind and current state to automaton identifiers (\texttt{automata} cell), and values
     to variables.
    
     The only time related feature available to \FLang{} execution machinery (in addition to packet arrival time) are interrupts. The configuration contains an \texttt{interrupts} cell. This cell contains a set of interrupt definitions (\texttt{interrupt*}). An interrupt is a triplet composed of: the time when the interrupt is to be triggered, the code to be executed, and
     a ``Time'' value equal to the interruption period for a periodic interruption or nothing for a non-periodic interruption.
     In addition, the \texttt{interrupts} cell contains an ordered list of the next ``times'' when an interrupt is to be executed.
    
     The \texttt{input} cell contains the current packet to be filtered, with its arrival time and port.
    %
     The configuration also contains a \texttt{k} cell that holds the \FLang{} statement under execution. Each time a new packet is input, the content of the \texttt{k} cell is replaced by the content of the \texttt{filter} cell.
    
     Finally, the \texttt{streams} cell contains: the
     packet input stream divided into the next packet to arrive (\texttt{inHead}) and the rest of the stream (\texttt{inTail}), the packet output stream, and the alarm output stream.
     In the input stream, resp. output stream, packets arriving, resp. leaving, on both ports are mixed together, but contains information on the port of entry, resp. exit.
     Some choices made to represent those streams are not an intrinsic part of the formal specification of \FLang{}. The division of the input stream into a head and a tail is such a choice. Those choices are made in order to be able to execute the specification. It is then required to implement, in the \K{} framework, a mechanism to retrieve and parse strings describing packet sequences sent to the filter. In order to help distinguish between the formal specification of \FLang{} and the mechanisms put in place to execute it, whenever possible, implementation choices, such as the format of strings describing packets, are defined in another file which is loaded with the \texttt{require} instruction.
    
    \subsection{Semantics}
    
     The formal specification of \FLang{}'s semantics relies on two auxiliary specifications. The first one define specific data types and associated functions (Fig.~\ref{fig:gpfpl:k-src-file:data-types}). The second one defines auxiliary conversion functions between those data types and \texttt{String}
     (Fig.~\ref{fig:k-src-file:string-conversion}).
    \begin{figure}[!htbp]
      \lstinputlisting[style=K]{./K-src/GPFPL/dataDefs-cleaned.k3}
      \caption{\K{} source file of specific data types}
      \label{fig:gpfpl:k-src-file:data-types}
    \end{figure} 
    \begin{figure}[!htbp]
      \lstinputlisting[style=K]{./K-src/GPFPL/stringConversions-cleaned.k3}
      \caption{\K{} source file of \texttt{String} conversion functions}
      \label{fig:k-src-file:string-conversion}
    \end{figure}
    
     The formal specification of \FLang{}'s semantics includes the usual rules for handling expressions that can be found in many \K{} examples or tutorials. The \texttt{strict} attributes of the syntax rules
     on lines~\ref{K:GPFPL:syntax:unaryOp} and~\ref{K:GPFPL:syntax:binaryOp}
     of Fig.~\ref{fig:gpfpl:k-src-file:syntax} specify that operation arguments in expressions have to be evaluated to values first.
    The rules in Fig.~\ref{fig:gpfpl:semantics:variables} specify the semantics of variables, which consists simply in retrieving their values in the corresponding configuration cell.
    \begin{figure}[!htbp]
    
      \begin{kSemDef}
        \input{./K-src/GPFPL/GPFPL-SEMANTICS-variablesSem.k.tex}
      \end{kSemDef}
    
      \caption{\FLang{}'s semantics for variables}
      \label{fig:gpfpl:semantics:variables}
    \end{figure}
     The \K{} source provided in Fig.~\ref{fig:gpfpl:k-src-file:expression-semantics} specifies the semantics of operations applied to values.
    \begin{figure}[!htbp]
      \lstinputlisting[style=K]{./K-src/GPFPL/gpfpl-expressionsSemantics-cleaned.k3}
      \caption{\K{} source file of \FLang{} expressions semantics}
      \label{fig:gpfpl:k-src-file:expression-semantics}
    \end{figure}

     The rest of \FLang{}'s semantics is decomposed in three execution phases:
    \begin{enumerate*}[label={(\alph*)},font={\bfseries}]
    \item \label{item:exec-phase:init} the program initialization,
    \item \label{item:exec-phase:select} the selection of the next statement to execute, and
    \item \label{item:exec-phase:exec} the execution of the selected statement.
    \end{enumerate*}
     Phase \ref{item:exec-phase:init} occurs only once at the beginning of the execution; then phases \ref{item:exec-phase:select} and \ref{item:exec-phase:exec} are repeatedly executed one after the other. Phase \ref{item:exec-phase:select} selects the statement associated to the next thing to do, i.e. filter a packet or execute an interruption. Phase \ref{item:exec-phase:exec} executes the selected statement.
    
    \subsubsection{Program Initialization Phase.}
    
     As specified by the rules in Fig.~\ref{fig:gpfpl:semantics:program}, the execution of a \FLang{} program is initialized by splitting the program in two. The prologue goes into the \texttt{prg} cell and the filter statement goes into the \texttt{filter} cell. Then the prologue elements are executed one by one.
    \begin{figure}[!htbp]
    
      \begin{kSemDef}
        \input{./K-src/GPFPL/GPFPL-SEMANTICS-programSplitSem.k.tex}
      \end{kSemDef}
    
      \caption{\FLang{} program top-level semantics in \K{}}
      \label{fig:gpfpl:semantics:program}
    \end{figure}
    
    
    
    
     The semantics of \texttt{AUTOMATON} prologues (Fig.~\ref{fig:gpfpl:semantics:prologue-automaton})
     is to create a new \texttt{automataKindDef} cell containing the definition of the automata kind.
    \begin{figure}[!htbp]
    
      \begin{kSemDef}
        \input{./K-src/GPFPL/GPFPL-SEMANTICS-prologAutomatonSem.k.tex}
      \end{kSemDef}
    
      \caption{Automaton prologue semantics}
      \label{fig:gpfpl:semantics:prologue-automaton}
    \end{figure}
    
     And, as specified by the rule in Fig.~\ref{fig:gpfpl:semantics:prologue-init}, the semantics of \texttt{INIT} prologues is to execute the associated statements. Any statement put in the \texttt{k} cell is to be executed, as specified in the remaining of this section.
    \begin{figure}[!htbp]
    
      \begin{kSemDef}
        \input{./K-src/GPFPL/GPFPL-SEMANTICS-prologInitSem.k.tex}
      \end{kSemDef}
    
      \caption{Initialization statement semantics}
      \label{fig:gpfpl:semantics:prologue-init}
    \end{figure}

    \subsubsection{Statement Selection Phase.}
    
     Once the prologue as been executed, and after every execution of an interruption or after filtering a packet, the semantics of \FLang{} is to select the next statement to be executed: either the filtering statement (in the \texttt{filter} cell), or the statement associated with the next interruption if this interruption is triggered before the next packet arrives.
    
     The process deciding which statement to execute next is specified in Fig.~\ref{fig:gpfpl:semantics:decideWhichStmtToLoad}. This process starts by loading (from the \texttt{inTail} stream cell) and parsing the next packet to filter while resetting the \texttt{time}, \texttt{port} and \texttt{fields} cells. Then, depending on which event happens first between the arrival of a new packet and the triggering of the next interruption, one of two helper commands (\texttt{loadInterrupt} or \texttt{loadNextPkt}) is put into the \texttt{k} cell to indicate which action is to be taken next.
    \begin{figure}[!htbp]
    
      \begin{kSemDef}
        \input{./K-src/GPFPL/GPFPL-SEMANTICS-decideWhichStmtToLoadSem.k.tex}
      \end{kSemDef}
    
      \caption{Semantic rules deciding which statement to execute next}
      \label{fig:gpfpl:semantics:decideWhichStmtToLoad}
    \end{figure}
     It is to be noted that this process is not intrinsically part of the specification of \FLang{}'s semantics. For practical reasons, this specification ``pre-loads'',  into the \texttt{inHead} cell, future packets that have not arrived yet. An implementation of \FLang{} would not ``pre-load'' packets; for an implementation, the input stream (concatenation of the \texttt{inHead} and \texttt{inTail}) is a whole. This fact is reflected by the fact that all the rules of Fig.~\ref{fig:gpfpl:semantics:decideWhichStmtToLoad} are structural.
    
     The formally specified semantics of \texttt{loadInterrupt} is shown in Fig.~\ref{fig:gpfpl:semantics:loadInterrupt}. The \texttt{loadInterrupt} command in the \texttt{k} cell is replaced by the statement associated with one of the interruptions scheduled to be triggered next (there can be many interruptions scheduled to be triggered at the same time). If the interruption triggered ($I$) is a recurring interruption, then $I$ is scheduled to be triggered again at $T + P$ where $T$ is the current time and $P$ is the period of the recurring interruption $I$. Otherwise, the \texttt{interrupt} cell of $I$ is simply removed.
    \begin{figure}[!htbp]
    
      \begin{kSemDef}
        \input{./K-src/GPFPL/GPFPL-SEMANTICS-loadInterruptSem.k.tex}
      \end{kSemDef}
    
      \caption{Semantics of \texttt{loadInterrupt}}
      \label{fig:gpfpl:semantics:loadInterrupt}
    \end{figure}
    
     Fig.~\ref{fig:gpfpl:semantics:loadNextPkt} graphically displays the formal semantics of \texttt{loadNextPkt}. Every field of the packet $P$ in the head of the input stream is loaded into the map of the \texttt{fields} environment cell. The \texttt{time} and \texttt{port} cells are set to the corresponding values associated to $P$. Finally, $P$ is removed from the head of the input stream, and the filtering statement in the \texttt{filter} cell is loaded into the \texttt{k} cell in replacement of the \texttt{loadNextPkt} command.
    \begin{figure}[!htbp]
    
      \begin{kSemDef}
        \input{./K-src/GPFPL/GPFPL-SEMANTICS-loadNextPktSem.k.tex}
      \end{kSemDef}
    
      \caption{Semantics of \texttt{loadNextPkt}}
      \label{fig:gpfpl:semantics:loadNextPkt}
    \end{figure}

    \subsubsection{Statement Execution Phase.}
    
     The definition of the \texttt{newInterrupt} statement semantics (Fig.~\ref{fig:gpfpl:semantics:newInterrupt-stmt}) uses one helper function, named \texttt{insertIntoOrderedTimeList}, which inserts an integer into an ordered list (Fig.~\ref{fig:gpfpl:semantics:newInterrupt-helpers}).
    \begin{figure}[!htbp]
    
      \begin{kSemDef}
        \input{./K-src/GPFPL/GPFPL-SEMANTICS-newInterruptSemHelpers.k.tex}
      \end{kSemDef}
    
      \caption{\FLang{}'s \texttt{newInterrupt} helper semantics}
      \label{fig:gpfpl:semantics:newInterrupt-helpers}
    \end{figure}
     As specified by the rules of Fig.~\ref{fig:gpfpl:semantics:newInterrupt-stmt}, the semantics of a \texttt{newInterrupt} statement is simply to create a new interruption in a new \texttt{interrupt} cell and to register this new interruption in the \texttt{nextInterrupts} cell.
    \begin{figure}[!htbp]
    
      \begin{kSemDef}
        \input{./K-src/GPFPL/GPFPL-SEMANTICS-newInterruptSem.k.tex}
      \end{kSemDef}
    
      \caption{\FLang{}'s \texttt{newInterrupt} statement semantics}
      \label{fig:gpfpl:semantics:newInterrupt-stmt}
    \end{figure}
    
     \FLang{}'s other statements semantics (specified in Fig.~\ref{fig:gpfpl:semantics:other-statements}) is quite simple. To execute a pair of statements, the first statement is executed and then the second one. The \texttt{strict} attribute of the conditional statement (\texttt{cond}) syntax rule (Fig.~\ref{fig:gpfpl:k-src-file:syntax}) specifies that the guard of the statement must be evaluated to a value first; then the rules in Fig.~\ref{fig:gpfpl:semantics:other-statements} specify that the sub-statement is executed only if the guard is true. Similarly, the \texttt{strict} attribute of the iteration statement syntax rule specifies that the controlling expression must be evaluated to a value first. If this controlling value is 0 then the execution of the iteration statement is over; otherwise its sub-statement is executed once and the iteration statement is executed again with its controlling expression decreased by 1.
    \begin{figure}[!htbp]
    
      \begin{kSemDef}
        \input{./K-src/GPFPL/GPFPL-SEMANTICS-otherStmtsSem.k.tex}
      \end{kSemDef}
    
      \caption{\FLang{}'s other statements semantics}
      \label{fig:gpfpl:semantics:other-statements}
    \end{figure}
    
     The semantics of the variable assignment command is quite standard (Fig.~\ref{fig:gpfpl:semantics:set-var-cmd}). The value associated to the variable in the map of the environment cell \texttt{vars} is updated to the new value of the variable. If the variable is not already present in the map of the \texttt{vars} cell, a structural rule adds it to the map, thus allowing the previous rule to apply.
    \begin{figure}[!htbp]
    
      \begin{kSemDef}
        \input{./K-src/GPFPL/GPFPL-SEMANTICS-setCmdSem.k.tex}
      \end{kSemDef}
    
      \caption{\FLang{}'s \texttt{set} command semantics}
      \label{fig:gpfpl:semantics:set-var-cmd}
    \end{figure}
    
     The semantics of automata-related commands is given in Fig.~\ref{fig:gpfpl:semantics:automata-cmds}. The \texttt{new\-Auto\-maton} command ``creates'' an new automaton of kind \texttt{K} and associates it to the variable \texttt{X}. The maps of the automata cell are updated to associate the kind \texttt{K} to the automaton referenced by \texttt{X}, and
     associate to \texttt{X} the initial state of automata of kind \texttt{K}. The \texttt{step} command sends the event \texttt{E} to the automaton referenced by \texttt{X}. If a transition triggered by \texttt{E} exists from the current state of the automaton, then the current state associated to \texttt{X} in the map of the \texttt{states} cell is updated with the new state; otherwise the error sub-statement \texttt{S} is executed.
    \begin{figure}[!htbp]
    
      \begin{kSemDef}
        \input{./K-src/GPFPL/GPFPL-SEMANTICS-automataCmdsSem.k.tex}
      \end{kSemDef}
    
      \caption{Automata commands semantics}
      \label{fig:gpfpl:semantics:automata-cmds}
    \end{figure}
    
    
    
    
     The \texttt{alarm} command semantics is provided in Fig.~\ref{fig:gpfpl:semantics:alarm-cmds}. Its semantics is simply to generate a packet on the alarm output stream.
    \begin{figure}[!htbp]
    
      \begin{kSemDef}
        \input{./K-src/GPFPL/GPFPL-SEMANTICS-alarmCmdsSem.k.tex}
      \end{kSemDef}
    
      \caption{Alert commands semantics}
      \label{fig:gpfpl:semantics:alarm-cmds}
    \end{figure}
    
     The packet related commands semantics (Fig.~\ref{fig:gpfpl:semantics:pkt-treatment-cmds}) relies on two internal commands: iSend, which sends a packet on the output stream; and iHalt, which halt the filtering process for the current packet.
     The \texttt{accept} command outputs the current packet and terminates the execution of the filter. The \texttt{drop} command terminates the execution of the filter.
     And \texttt{send} outputs a packet.
    \begin{figure}[!htbp]
    
      \begin{kSemDef}
        \input{./K-src/GPFPL/GPFPL-SEMANTICS-decisionCmdsSem.k.tex}
      \end{kSemDef}
    
      \caption{packet-related commands semantics}
      \label{fig:gpfpl:semantics:pkt-treatment-cmds}
    \end{figure}
    
\end{FLangFullSpec}

\section{Testing \FLang{}'s Specification}
\label{sec:testing}
    
 {\FLang}'s specification, introduced above and contained in the companion technical report~\cite{le-guernic:2016:erfspflukf}, is not necessarily perfect. By a matter of fact, imperfections of {\FLang}'s specification are of interest to the experimentation reported in this paper. Indeed, the goal of the experimentation is to see how a tool such as the {\K} framework can help to spot and correct imperfections in filtering language specifications. One way to do so is by ``testing'' the new language specified, which is possible if the framework used to specify the language supports the execution or simulation of language specifications, which is the case for the {\K} framework.

 The test scenario used assumes a network of clients and servers. The clients request resources to servers using a made-up protocol, called ``DHCP cherry'', summarized in Fig.~\ref{fig:DHCP-cherry:nominal-pkt-seq}.
\begin{figure}[!htb]
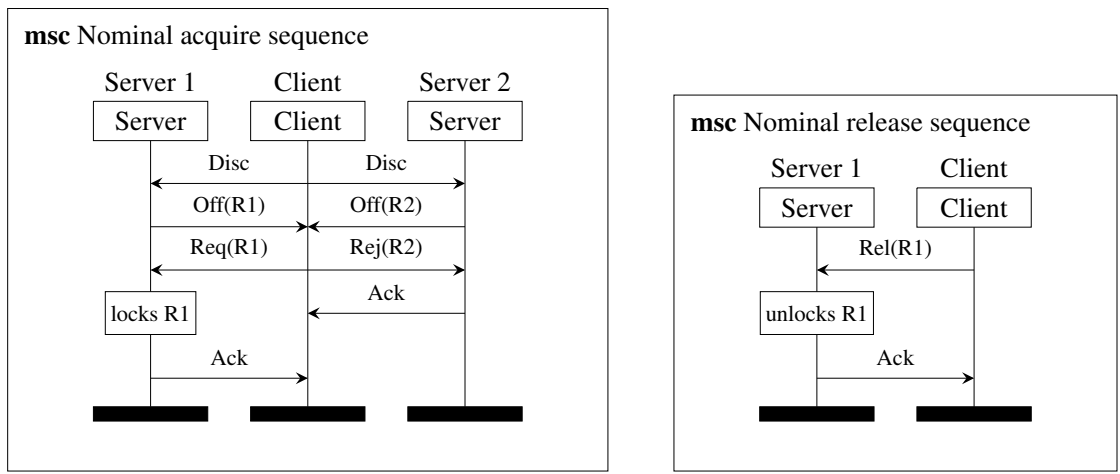

  
  \newcommand{\scalingFactor}{.95}%

  \hspace*{\stretch{1}}
  %
  \scalebox{\scalingFactor}{
    \begin{msc}{Nominal acquire sequence}
      \setlength{\levelheight}{3.5ex}
      \declinst{S1}{Server 1}{Server}
      \declinst{Cl}{Client}{Client}
      \declinst{S2}{Server 2}{Server}
      
      \mess{\smaller Disc}{Cl}{S1}
      \mess{\smaller Disc}{Cl}{S2}
      \nextlevel
      \mess{\smaller Off(R1)}{S1}{Cl}
      \mess{\smaller Off(R2)}{S2}{Cl}
      \nextlevel
      \mess{\smaller Req(R1)}{Cl}{S1}
      \mess{\smaller Rej(R2)}{Cl}{S2}
      \nextlevel[.5]
      \action*{\smaller locks R1}{S1}
      \nextlevel[.5]
      \mess{\smaller Ack}{S2}{Cl}
      \nextlevel[1.5]
      \mess{\smaller Ack}{S1}{Cl}
    \end{msc}
  }
  \hspace*{\stretch{1}}
  %
  \scalebox{\scalingFactor}{
    \begin{msc}{Nominal release sequence}
      \setlength{\levelheight}{3.5ex}
      \declinst{S1}{Server 1}{Server}
      \declinst{Cl}{Client}{Client}
      
      \mess{\smaller Rel(R1)}{Cl}{S1}
      \nextlevel[.5]
      \action*{\smaller unlocks R1}{S1}
      \nextlevel[2]
      \mess{\smaller Ack}{S1}{Cl}
    \end{msc}
  }
  \hspace*{\stretch{1}}

  \caption{Nominal packet sequences of DHCP cherry protocol}
  \label{fig:DHCP-cherry:nominal-pkt-seq}
\end{figure}
 The test scenario assumes that servers behave poorly when interacting concurrently with different clients. The objective of the test scenario is then to filter communications in front of servers in order to prevent any concurrent client-server interactions with any given server.
 This test scenario is obviously made-up for this experimentation, which is a requirement due to confidentiality issues. However, it is still covering the most frequently used features of filtering languages similar to {\FLang}, while remaining simple enough for a first experimentation.

     From the point of view of servers, non-concurrent interactions are sequential instances of only three generic atomic packet sequences. Those atomic packet sequences are the ones accepted by the automaton in Fig.~\ref{fig:automaton-pkt-seq}.
     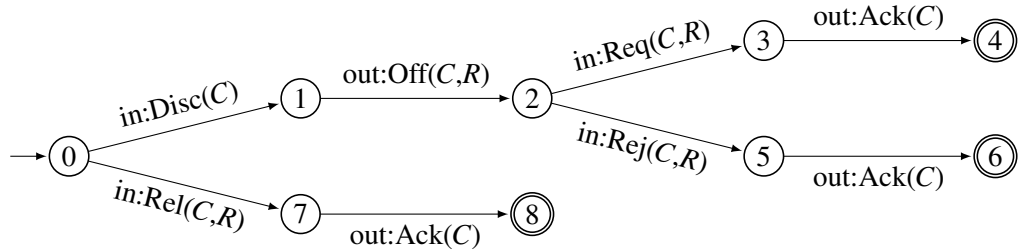
\begin{figure}[!htbp]
       \centering
       
       \begin{tikzpicture}[
           shorten >=1pt, -latex, >=latex, initial text={},
           coord/.style={inner sep=0pt, outer sep=0pt},
           myState/.style={draw, semithick,  circle, inner sep=2pt, outer sep=0pt},
         ] 
         \begin{scope}[node distance=8em, on grid, auto, font={\smaller[0]}]
           \def\vertSpacing{2em}
           \node[myState, initial] (sInit) {0};
           \node[coord] (coordNextSInit) [right=of sInit] {};
           \node[myState] (sDisc) [above=\vertSpacing of coordNextSInit] {1};
           \node[myState] (sOff) [right=of sDisc] {2};
           \node[coord] (coordNextSOff) [right=of sOff] {};
           \node[myState] (sReq) [above=\vertSpacing of coordNextSOff] {3};
           \node[myState, accepting] (sAckReq) [right=of sReq] {4};
           \node[myState] (sRej) [below=\vertSpacing of coordNextSOff] {5};
           \node[myState, accepting] (sAckRej) [right=of sRej] {6};
           \node[myState] (sRel) [below=\vertSpacing of coordNextSInit] {7};
           \node[myState, accepting] (sAckRel) [right=of sRel] {8};
         \end{scope}
         
         \begin{scope}[sloped, font={\smaller[0]}]
           \path (sInit) edge node[above]{in:Disc($C$)} (sDisc);
           \path (sDisc) edge node[above]{out:Off($C$,$R$)} (sOff);
           \path (sOff) edge node[above]{in:Req($C$,$R$)} (sReq);
           \path (sReq) edge node[above]{out:Ack($C$)} (sAckReq);
           \path (sOff) edge node[below]{in:Rej($C$,$R$)} (sRej);
           \path (sRej) edge node[below]{out:Ack($C$)} (sAckRej);
           \path (sInit) edge node[below]{in:Rel($C$,$R$)} (sRel);
           \path (sRel) edge node[below]{out:Ack($C$)} (sAckRel);
         \end{scope}
       \end{tikzpicture}
    
       \caption{Automaton of server-side atomic packet sequences}
       \label{fig:automaton-pkt-seq}
     \end{figure}
     In this automaton, ``in:$MP$'', resp. ``out:$MP$'', is a transition trigger matching any incoming packet (from the rest of the network to the server), resp. outgoing packet, matching packet pattern $MP$. $C$, resp. $R$, is a client, resp. ressource, identifier variable. $C$, resp. $R$, has to be instantiated in the same way (have the same value) for any packet of the same atomic packet sequence accepted by the automaton.
 The automaton of Fig.~\ref{fig:automaton-pkt-seq} is refined into a filtering policy automaton described in Fig.~\ref{fig:filtering-policy-automaton}. Variables $C$ and $R$ have the same constraints as for the automaton of Fig.~\ref{fig:automaton-pkt-seq}. The variable ``$\ast$'' matches any value, packet pattern ``out:$\ast$'' matches any outgoing packet, and packet pattern ``out:$\ast$ - Ack($C$)'' matches any outgoing packet except Ack($C$).
 \begin{figure}[!htbp]
   \centering
   
   \begin{tikzpicture}[
       shorten >=1pt, -latex, >=latex, initial text={},
       coord/.style={inner sep=0pt, outer sep=0pt},
       myState/.style={draw, semithick,  circle, inner sep=2pt, outer sep=0pt},
     ] 
     \begin{scope}[node distance=8em, on grid, auto, font={\smaller[0]}]
       \def\vertSpacing{2em}
       \node[myState, initial above, accepting] (sInit) {0};
       \node[myState] (sDisc) [right=of sInit] {1};
       \node[myState] (sRe) [right=of sDisc] {2};
       \node[myState] (sRel) [left=of sInit] {3};
     \end{scope}
     
     \begin{scope}[sloped, font={\smaller[0]}]
       \path (sInit) edge[loop below] node[below]{out:$\ast$} ();
       \path (sInit) edge node[above]{in:Disc($C$)} (sDisc);
       \path (sDisc) edge[loop above] node[above]{out:$\ast$} ();
       \path (sDisc) edge[bend left] node[pos=0.6,above,sloped=false]{in:Req($C$,$\ast$)} (sRe);
       \path (sDisc) edge node[below]{in:Rej($C$,$\ast$)} (sRe);
       \path (sRe) edge[loop right] node[sloped=false,align=center]{out:$\ast$ - \\ Ack($C$)} ();
       \path (sRe) edge[bend left,out=50] node[pos=0.45,below,sloped=false]{out:Ack($C$)} (sInit);
       \path (sInit) edge node[above]{in:Rel($C$,$\ast$)} (sRel);
       \path (sRel) edge[loop left] node[sloped=false,align=center]{out:$\ast$ - \\ Ack($C$)} ();
       \path (sRel) edge[bend right] node[below]{out:Ack($C$)} (sInit);
     \end{scope}
   \end{tikzpicture}

   \caption{Filtering policy automaton}
   \label{fig:filtering-policy-automaton}
 \end{figure}
 This filtering policy accepts every outgoing packet; thus having no effect on the packets generated by the server. For incoming packets, if the current state of the automaton has no transition whose trigger matches the packet then the packet is discarded; otherwise, the packet is accepted and the associated transition is triggered.
 This filtering policy assumes that clients comply with the DHCP cherry protocol and ensures only that the filtered server only interacts sequentially with clients. If there is no idle server ready to receive a packet from a client, this client gets no answer and is expected to retry later.

 This policy has been encoded in {\FLang} and executed using the following command (in Linux Bash):
 ``{\ttfamily krun dhcp.gpfpl < dhcp\_input-dataset.txt > dhcp\_output.txt}''
 where the file {\ttfamily dhcp\_in\-put-data\-set.txt} contains a sequence of packets already ``parsed'' (decoded packets, Fig.~\ref{fig:filter-architecture}) input to the filter. The output of the simulation of the code (\texttt{dhcp.gpfpl}) written in the specified language ({\FLang}) is written in \texttt{dhcp\_output.txt}.

\begin{FLangFullTest}
    \section{Testing \FLang{}'s Specification}
    \label{sec:testing}
    
     The above specification of \FLang{} syntax and semantics is not necessarily perfect. By a matter of fact, the imperfections of \FLang{}'s specification are of interest to the experimentation reported in this paper. Indeed, the goal of the experimentation is to see how a tool such as the \K{} framework can help to spot imperfections in filtering language specifications, and help correct them. One way to do so, is by ``testing'' the new language specified, which is possible if the framework used to specify the language supports the execution or simulation of language specifications, which is the case for the \K{} framework.
    
     The test scenario used assumes a network of clients and servers. The clients request resources to servers using a made-up protocol called ``DHCP cherry''. The test scenario assumes that servers behave poorly when interacting concurrently with different clients. The objective of the test scenario is then to filter communications towards servers, as architectured in Fig.~\ref{fig:test-scenario-architectue}, in order to prevent any concurrent client-server interactions with any given server.
    
    \begin{figure}[!htbp]
      \centering
       
      \newcommand{\clientIcon}{256x256_devices_computer}
      \newcommand{\serverIcon}{256x256_places_network-server}
    
      \begin{tikzpicture}[
          coord/.style={inner sep=0pt, outer sep=0pt},
          icon/.style={inner sep=0pt, outer sep=0pt},
          concNetwork/.style={ultra thick},
          seqNetwork/.style={densely dotted, thick},
          baseline=(current bounding box.center)
        ]
    
        \begin{scope}[node distance=5em]
          \node[coord] (NetPlug_client1) {};
          \node[coord] (NetPlug_client2) [right=of NetPlug_client1] {};
          \node[coord] (NetPlug_server1) [right=of NetPlug_client2] {};
          \node[coord] (NetPlug_client3) [right=of NetPlug_server1] {};
          \node[coord] (NetPlug_server2) [right=of NetPlug_client3] {};
          \node[coord] (NetPlug_client4) [right=of NetPlug_server2] {};
        \end{scope}
        \begin{scope}[node distance=.5em]
          \newcommand{\clientIconWidth}{3em}
          \newcommand{\filterIconWidth}{3em}
          \newcommand{\serverIconWidth}{3em}
          \node[icon] (client1) [above=of NetPlug_client1] {\includegraphics[width=\clientIconWidth]{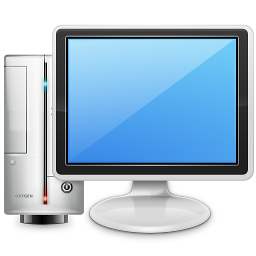}};
          \node[icon] (client2) [above=of NetPlug_client2] {\includegraphics[width=\clientIconWidth]{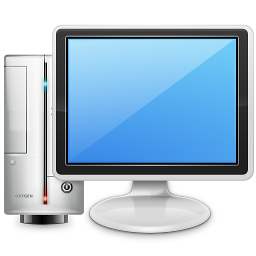}};
          \node[icon] (filter1) [above=of NetPlug_server1] {\includegraphics[width=\clientIconWidth]{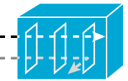}};
          \node[icon] (server1) [above=of filter1] {\includegraphics[width=\serverIconWidth]{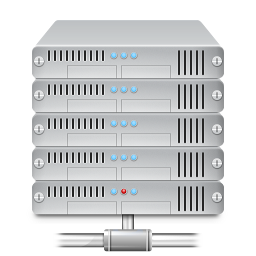}};
          \node[icon] (client3) [above=of NetPlug_client3] {\includegraphics[width=\clientIconWidth]{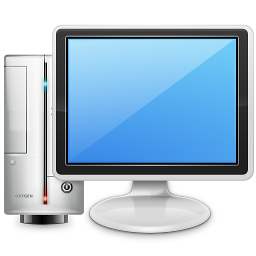}};
          \node[icon] (filter2) [above=of NetPlug_server2] {\includegraphics[width=\clientIconWidth]{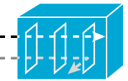}};
          \node[icon] (server2) [above=of filter2] {\includegraphics[width=\serverIconWidth]{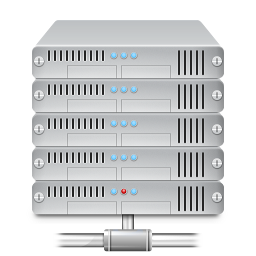}};
          \node[icon] (client4) [above=of NetPlug_client4] {\includegraphics[width=\clientIconWidth]{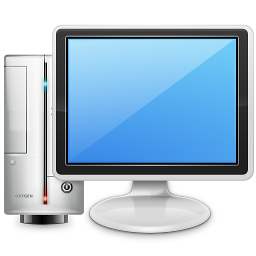}};
        \end{scope}
          
        \coordinate (nwLeft) at ([xshift=-1em] NetPlug_client1);
        \coordinate (nwRight) at ([xshift=1em] NetPlug_client4);
        \begin{scope}[concNetwork]
          \draw (nwLeft) -- (nwRight);
          \draw (NetPlug_client1) -- (client1);
          \draw (NetPlug_client2) -- (client2);
          \draw (NetPlug_server1) -- (filter1);
          \draw (NetPlug_client3) -- (client3);
          \draw (NetPlug_server2) -- (filter2);
          \draw (NetPlug_client4) -- (client4);
        \end{scope}
        \begin{scope}[seqNetwork]
          \draw (filter1) -- (server1);
          \draw (filter2) -- (server2);
        \end{scope}
          
      \end{tikzpicture}
    
      \caption{Network architecture of the test scenario}
      \label{fig:test-scenario-architectue}
    \end{figure}
    
     This test scenario is obviously made-up for this experimentation, which is a requirement due to confidentiality issues. However, it is still covering the most frequently used features of filtering languages similar to \FLang{}, while remaining simple enough for a sub-part of an experimentation.
    
    \subsection{DHCP cherry}
    
     The protocol used for this test scenario is a simplified version of the DHCP protocol. Packet formats and nominal sequences are described below.
    
    \subsubsection{Protocol}
    
     The protocol for DHCP cherry follows the packet sequences described in Fig.~\ref{fig:DHCP-cherry:nominal-pkt-seq}.
    \begin{figure}[!htp]
      
      \newcommand{\scalingFactor}{.645}%
    
      %
      \scalebox{\scalingFactor}{%
        \begin{hmsc}{Global sequencing}(-1.4,-4)(1.4,2)
          \hmscstartsymbol{start}(0,0)
          \hmscreference{acq}{Acquire}(0,-1)
          \hmscreference{rel}{Release}(0,-2)
          \hmscendsymbol{end}(0,-3)
          \arrow{start}{acq}
          \arrow{acq}{rel}
          \arrow{rel}{end}
        \end{hmsc}
      }
      \hspace*{\stretch{1}}
      %
      \scalebox{\scalingFactor}{
        \begin{msc}{Nominal acquire sequence}
          \setlength{\levelheight}{3.5ex}
          \declinst{S1}{Server 1}{Server}
          \declinst{Cl}{Client}{Client}
          \declinst{S2}{Server 2}{Server}
          
          \mess{\smaller Disc}{Cl}{S1}
          \mess{\smaller Disc}{Cl}{S2}
          \nextlevel
          \mess{\smaller Off(R1)}{S1}{Cl}
          \mess{\smaller Off(R2)}{S2}{Cl}
          \nextlevel
          \mess{\smaller Req(R1)}{Cl}{S1}
          \mess{\smaller Rej(R2)}{Cl}{S2}
          \nextlevel[.5]
          \action*{\smaller locks R1}{S1}
          \nextlevel[.5]
          \mess{\smaller Ack}{S2}{Cl}
          \nextlevel[1.5]
          \mess{\smaller Ack}{S1}{Cl}
        \end{msc}
      }
      \hspace*{\stretch{1}}
      %
      \scalebox{\scalingFactor}{
        \begin{msc}{Nominal release sequence}
          \setlength{\levelheight}{3.5ex}
          \declinst{S1}{Server 1}{Server}
          \declinst{Cl}{Client}{Client}
          
          \mess{\smaller Rel(R1)}{Cl}{S1}
          \nextlevel[.5]
          \action*{\smaller unlocks R1}{S1}
          \nextlevel[2]
          \mess{\smaller Ack}{S1}{Cl}
        \end{msc}
      }
      %
    
      \caption{Nominal packet sequences of DHCP cherry protocol}
      \label{fig:DHCP-cherry:nominal-pkt-seq}
    \end{figure}
     The client starts by broadcasting a request for resource (\texttt{Discover} packet). Servers answer with resource offers (\textsf{Offer} packet), but do not lock the resource for the client yet. The client chooses one of the offered resources ($R1$) and sends a request for that resource (\textsf{Request} packet) and rejections (\textsf{Reject} packet) for the other resources. Servers which received a rejection packet then send an acknowledgment packet (\textsf{Acknowledge} packet). The server, which received a request packet, locks the associated resource for the client and sends him an acknowledgment. The client is then free to use the resource for as long as he wishes. Once done with the resource, the client releases the resource to the server (\textsf{Release} packet). And the server acknowledges reception of the release packet.
    
     A client that does not receive any offer to a discovery request, or an expected acknowledgment, is supposed to try again later to emit the packet to which it did not receive an answer. However, there is no explicit recovery mechanism in the protocol. If a packet sequence stops between the \textsf{Request} packet and the \textsf{Release} packet, the associated resource is ``lost''.
    
    \subsubsection{Packet formats}
    
     The format of packets is given in Fig.~\ref{fig:dhcp-cherry:pkt-fmt}.
    \begin{figure}[!htbp]
      \centering
    
      \begin{tikzpicture}[remember picture]
        \begin{scope}[rectangle, minimum height=4ex]
          \node[draw](protoPktType) {Pkt Type};
          \node[draw, anchor=west](client) at (protoPktType.east) {Client Id};
          \node[draw, dashed, anchor=west](resource) at (client.east) {Resource Id};
        \end{scope}
        \begin{scope}[font={\smaller[3]}, inner sep=0.2ex, outer sep=0pt]
          \node[anchor=south west] at (protoPktType.north west) {0};
          \node[anchor=south east] at (protoPktType.north east) {3};
          \node[anchor=south west] at (client.north west) {4};
          \node[anchor=south east] at (client.north east) {7};
          \node[anchor=south west] at (resource.north west) {8};
          \node[anchor=south east] at (resource.north east) {11};
        \end{scope}
      \end{tikzpicture}
      \\[3ex]
      { 
        \begin{tabular}{|l|l|c|c|r|}
          \hline
          \hspace*{\stretch{1}} Packet \hspace*{\stretch{1}}
          & \hspace*{\stretch{1}} MSC \hspace*{\stretch{1}}
          & \begin{tikzpicture}[remember picture, baseline=(encoding.base)]
              \node (encoding) {Type encoding};
            \end{tikzpicture}
          & \begin{tikzpicture}[remember picture, baseline=(resourcePresente.base)]
              \node (resourcePresente) {Resource part};
            \end{tikzpicture}
          & \hspace*{\stretch{1}} Pkt length \hspace*{\stretch{1}} \\
          \hline
          Discover    & Disc   & 0 & NO &  8 bits \\
          Offer       & Off(R) & 1 & YES & 12 bits \\
          Request     & Req(R) & 2 & YES & 12 bits \\
          Reject      & Rej(R) & 3 & YES & 12 bits \\
          Release     & Rel(R) & 4 & YES & 12 bits \\
          Acknowledge & Ack    & 5 & NO &  8 bits \\
          \hline
        \end{tabular}
        \begin{tikzpicture}[remember picture, overlay, -latex, draw=black!50, shorten <=2pt, shorten >=2pt]
          \draw (protoPktType) to[out=-70,in=110] (encoding);
          \draw (resource) to (resourcePresente);
        \end{tikzpicture}
      }
    
      \caption{Format of DHCP cherry packets}
      \label{fig:dhcp-cherry:pkt-fmt}
    \end{figure}
     A packet is 8 or 12 bits long. A packet starts by a 4 bits long packet type identifier (Discover, Offer, \dots), followed by a 4 bits long client identifier identifying the client involved in the session. If the packet carries a resource identifier, a 4 bits long resource identifier is appended at the end of the packet.
    
    \subsection{The Filtering Policy to Enforce}
    
     From the point of view of servers, non-concurrent interactions are sequential instances of only three generic atomic packet sequences. Those atomic packet sequences are the ones accepted by the automaton in Fig.~\ref{fig:automaton-pkt-seq}.
     \begin{figure}[!htbp]
       \centering
       
       \begin{tikzpicture}[
           shorten >=1pt, -latex, >=latex, initial text={},
           coord/.style={inner sep=0pt, outer sep=0pt},
           myState/.style={draw, semithick,  circle, inner sep=2pt, outer sep=0pt},
         ] 
         \begin{scope}[node distance=8em, on grid, auto, font={\smaller[0]}]
           \def\vertSpacing{2em}
           \node[myState, initial] (sInit) {0};
           \node[coord] (coordNextSInit) [right=of sInit] {};
           \node[myState] (sDisc) [above=\vertSpacing of coordNextSInit] {1};
           \node[myState] (sOff) [right=of sDisc] {2};
           \node[coord] (coordNextSOff) [right=of sOff] {};
           \node[myState] (sReq) [above=\vertSpacing of coordNextSOff] {3};
           \node[myState, accepting] (sAckReq) [right=of sReq] {4};
           \node[myState] (sRej) [below=\vertSpacing of coordNextSOff] {5};
           \node[myState, accepting] (sAckRej) [right=of sRej] {6};
           \node[myState] (sRel) [below=\vertSpacing of coordNextSInit] {7};
           \node[myState, accepting] (sAckRel) [right=of sRel] {8};
         \end{scope}
         
         \begin{scope}[sloped, font={\smaller[1]}]
           \path (sInit) edge node[above]{in:Disc($C$)} (sDisc);
           \path (sDisc) edge node[above]{out:Off($C$,$R$)} (sOff);
           \path (sOff) edge node[above]{in:Req($C$,$R$)} (sReq);
           \path (sReq) edge node[above]{out:Ack($C$)} (sAckReq);
           \path (sOff) edge node[below]{in:Rej($C$,$R$)} (sRej);
           \path (sRej) edge node[below]{out:Ack($C$)} (sAckRej);
           \path (sInit) edge node[below]{in:Rel($C$,$R$)} (sRel);
           \path (sRel) edge node[below]{out:Ack($C$)} (sAckRel);
         \end{scope}
       \end{tikzpicture}
    
       \caption{Automaton of Server-side Atomic Packet Sequences}
       \label{fig:automaton-pkt-seq}
     \end{figure}
     In this automaton, ``in:$MP$'', resp. ``out:$MP$'', is a transition trigger matching any incoming packet (from the rest of the network to the server), resp. outgoing packet (from the server to the rest of the network), matching packet pattern $MP$. $C$ and $R$ are variables. $C$ is a client identifier variable. $R$ is a resource identifier variable. $C$, resp. $R$, has to be instantiated in the same way (have the same value) for any packet of the same atomic packet sequence accepted by the automaton.
    
     The automaton of Fig.~\ref{fig:automaton-pkt-seq} is refined into a filtering policy automaton described in Fig.~\ref{fig:filtering-policy-automaton}. Variables $C$ and $R$ have the same constraints as for the automaton of Fig.~\ref{fig:automaton-pkt-seq}. The variable ``$\ast$'' matches any value, packet pattern ``out:$\ast$'' matches any outgoing packet, and packet pattern ``out:$\ast$ - Ack($C$)'' matches any outgoing packet except Ack($C$).
     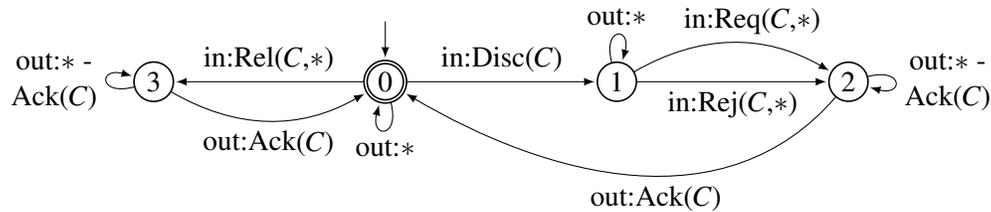
\begin{figure}[!htbp]
       \centering
       
       \begin{tikzpicture}[
           shorten >=1pt, -latex, >=latex, initial text={},
           coord/.style={inner sep=0pt, outer sep=0pt},
           myState/.style={draw, semithick,  circle, inner sep=2pt, outer sep=0pt},
         ] 
         \begin{scope}[node distance=8em, on grid, auto, font={\smaller[0]}]
           \def\vertSpacing{2em}
           \node[myState, initial above, accepting] (sInit) {0};
           \node[myState] (sDisc) [right=of sInit] {1};
           \node[myState] (sRe) [right=of sDisc] {2};
           \node[myState] (sRel) [left=of sInit] {3};
         \end{scope}
         
         \begin{scope}[sloped, font={\smaller[1]}]
           \path (sInit) edge[loop below] node[below]{out:$\ast$} ();
           \path (sInit) edge node[above]{in:Disc($C$)} (sDisc);
           \path (sDisc) edge[loop above] node[above]{out:$\ast$} ();
           \path (sDisc) edge[bend left] node[pos=0.6,above,sloped=false]{in:Req($C$,$\ast$)} (sRe);
           \path (sDisc) edge node[below]{in:Rej($C$,$\ast$)} (sRe);
           \path (sRe) edge[loop right] node[sloped=false,align=center]{out:$\ast$ - \\ Ack($C$)} ();
           \path (sRe) edge[bend left,out=50] node[pos=0.45,below,sloped=false]{out:Ack($C$)} (sInit);
           \path (sInit) edge node[above]{in:Rel($C$,$\ast$)} (sRel);
           \path (sRel) edge[loop left] node[sloped=false,align=center]{out:$\ast$ - \\ Ack($C$)} ();
           \path (sRel) edge[bend right] node[below]{out:Ack($C$)} (sInit);
         \end{scope}
       \end{tikzpicture}
    
       \caption{Filtering Policy Automaton}
       \label{fig:filtering-policy-automaton}
     \end{figure}
     This filtering policy accepts every outgoing packet; thus having no effect on the packets generated by the server. For incoming packets, if the current state of the automaton has no transition whose trigger matches the packet then the packet is discarded; otherwise, the packet is accepted and the associated transition is triggered.
    
     This filtering policy assumes that clients comply with the DHCP cherry protocol and ensures only that the filtered server only interacts sequentially with clients. If there is no idle server ready to receive a packet from a client, this client gets no answer and is expected to retry later.
    
    \subsection{The Filter Code in \FLang{}}
    \label{sec:flang-code-of-dhcp-filter}
    
     The \FLang{}'s code for this filtering policy is contained the file \texttt{dhcp.gpfpl}, displayed
     below.
    %
       \noindent\lstinputlisting[style=FLang]{./K-src/GPFPL/dhcp.gpfpl}
    %
     The states of the filtering automaton of Fig.~\ref{fig:filtering-policy-automaton} are directly encoded in an automata kind definition in the prologue, with generic triggering conditions that only encode the type of packet received. A unique instance (\texttt{\#A}) of this kind of automata is created.
     For every packet received by the filter, additional triggering conditions (packet input port, client identifier, {\dots}) are handled in the \texttt{FILTER} code itself.
     If a packet is received with a type compatible with additional triggering conditions, the packet type is sent to the automaton \texttt{\#A} to verify that the current state is compatible with the reception of this type of packet, and update the state of the automaton.
    
     In addition, every dropped packet increments a counter (\texttt{ignoredPktCnt}), which is reset to 0 every 60 ``time unit'' by a recurrent interruption initialized in the prologue. If this counter reaches the threshold (5), an alarm is raised to warn that ``many'' packets are dropped by the filter.
    
    \subsection{Simulating the Filter}
    
     The above filtering code written in \FLang{} can then be simulated by running the following command (in Linux Bash) :
     \begin{center}\ttfamily
       krun dhcp.gpfpl < dhcp\_input-dataset.txt > dhcp\_output.txt
     \end{center}
     where \texttt{dhcp\_input-dataset.txt} contains a sequence of ``parsed'' packets (decoded packets, Fig.~\ref{fig:filter-architecture}) input to the filter. The output of the simulation of the code (\texttt{dhcp.gpfpl}) written in the specified language ({\FLang}) is written in \texttt{dhcp\_output.txt}.
    
     The following input (\texttt{dhcp\_input-dataset.txt}):
       \noindent\lstinputlisting{./K-src/GPFPL/dhcp_input-dataset.txt}
    produces the following expected output (\texttt{dhcp\_output.txt}):
       \noindent\lstinputlisting[firstline=1,lastline=12]{./K-src/GPFPL/dhcp_output.txt}
    appended with a description of the final configuration (which is not displayed here).
\end{FLangFullTest}

\section{Discussion on the Experimentation}
\label{sec:retex}

 The primary goal of this paper is not to set out the filtering policy described in Sect.~\ref{sec:testing} or, even, {\FLang}'s specification described in Sect.~\ref{sec:flang-formal-specification}. This paper is an industrial experience report on
 a primary evaluation of the cost and benefits of using formal specification tools in general,
 and the {\K} framework in particular, to formally specify the syntax and semantics of filtering languages.
 Overall, it seems to the authors that using the {\K} framework helped greatly to improve \FLang{}'s specification quality. It forced the specification authors to be precise, and helped spot various errors and missing specification fragments.

 With regard to the ``cost'', this experimentation argues in favor of tool supported formal specifications for high quality specifications of filtering languages. Of course, using natural language, it is possible to produce a cheaper, but ambiguous and approximate, specification. However,
 from the authors' natural language based experiences with packet filtering language specifications,
 using natural language to produce a specification with a similar level of precision and correctness would be more costly for engineers with operational semantics knowledge. With a decent knowledge of operational semantics concepts, the cost for newcomers to the {\K} framework is relatively low, thanks to the numerous tutorials (in text and video), manuals and examples. In fact, having been exposed to operational semantics concepts (apart from general computer science concepts) seems to be the only prerequisite to efficiently using the {\K} framework.

 From the authors' previous experiences at formal specification of packet filtering language specifications without tool support, the cost of the constraints imposed by the {\K} framework seems to the authors to be lower than the benefits provided by the tool support. Typically, the ability to simulate\footnote{The authors prefer to talk of ``simulation'' rather than ``execution'', as the loading time of the execution environment and limited ability to interact with other components would most likely prevent to use such an execution in a real world setting.} the formal specification of the filtering language requires a particular handling of input/output related rules. However, this same ability to simulate the formal specification of the filtering language is highly beneficial when validating the correctness of the specification and expressivity of the language by ``executing'' test and documentation programs.

 Other benefits of tool supported formal specifications of languages are numerous. In natural language documents specifying new languages, it is too common for program examples to be inconsistent with the language grammar. It is easily explained by the modifications brought to the language grammar during the specification document development. Examples directly related to the modified statements are usually modified accordingly. However, examples related to other aspects of the language are often forgotten. Using a tool supported formal specification, it is easy to adopt a ``continuous/frequent integration'' approach where examples are: written in separate files, regularly parsed to verify that they comply with the current grammar, and automatically imported in the specification document (the creation of this paper used this approach).

 Additionally, use of a tool-supported formal specification approach modifies the workflow often applied when using natural language specification documents. With natural language specifications, the specification document writing process usually starts early after a short engineering phase (it may not be true for a language \emph{development} process, however it is often the case in pure language \emph{specification} processes), and the main part of the language specification is done during the specification document writing process. With a tool-supported formal specification approach, the specification of the language tend to be first developed inside the tool, and then the language specification is clarified during the specification document writing process. With a tool-supported formal specification approach, the language specification becomes a two phases process with two different views on the language specification. The ``two different views'' aspect is particularly true with the {\K} framework were semantics rules are entered textually in the source file and can be rendered graphically for the specification document. This two phases workflow (development then clarification and documentation) helps spot: differences of treatments (in particular for configuration cells), generalization and reuse opportunities (for example, in this experimentation, the use of only two internal commands, \texttt{iSend} and \texttt{iHalt}, to encode the three packet commands \texttt{accept}, \texttt{drop} and \texttt{send}), different concepts that are candidates to modularization (for example, in this experimentation, the externalization of packet data type definitions and string conversions), errors that manifest themselves in rare occasions (for example, in an earlier version of {\FLang}, automaton states and variable values where stored in the same map, which could trigger a key clash caused by variable and automaton identifiers having the same ``name'' part), or general simplifications (for example, during this report writing process, {\FLang}'s configuration has been heavily reformatted to simplify the language specification and be closer to the concepts manipulated). From the authors experience, in general and compared to a natural language approach, a tool-supported formal specification process helps simplify and clarify a language specification.




 Moreover, the ability to execute the formal specification allows to adopt an incremental approach for the specification of the different statements semantics. In such an approach, the syntax of the language is first specified. Then a program example making use of all the statements of the language in as much context as reasonable is written. The semantics of the statements is then defined statements by statements. The program is executed using {\K}'s run time; and the execution stops when reaching a statement whose semantics is not defined yet. All the semantics rules associated to this statement are then defined. When stopping an execution, {\K}'s run time displays the current state of the configuration which can help specify the missing semantics rules.
 As the test program execution goes further and further during the language semantics specification process, this incremental approach is more rewarding for people in charge of the specification. The impact of using this incremental approach (which is not required by the {\K} framework) on the quality of the specifications produced remains to be investigated.

 Finally, the ability to execute the formal specification allows to test and validate the language specification. Two important points to validate are: the expressivity of the language and its expected semantics.
 {\FLang}'s test code (Sect.~\ref{sec:testing}) provided in the companion technical report~\cite{le-guernic:2016:erfspflukf} emphasizes the limitations of the simple automata that can be defined using {\FLang}. It could be useful to have automaton state variables, and triggering conditions that test and check automaton state variable values. However, adding automaton state variables would complexify automata definitions.
 Similarly, {\FLang}'s test code contains a recurring code sequence to handle alarms which is triggered only when a threshold of a specific event occurrences is reached. It could be useful to add a specific command to {\FLang} which would have the same semantics as this recurring sequence.
 The ability to test programs does not solve expressivity questions (which have to be answered on a per language basis), however it helps explicit those questions.
 With regard to expected semantics, writing test programs helps validate that programs have the semantics that users would expect. The initial version of {\FLang}'s test code did not behave as expected. It ended up being a misplaced statement in the filter code, but could also have been a problem with the semantics specification.
 Discovering the cause of a misbehavior of a test program (error in the semantics or the program) could be greatly simplified by {\K}'s debugger which can ``execute'' formal specifications step by step; especially as Domain Specific Languages (specifications and implementations) usually have limited debugging facilities (which is in accordance with their philosophy of limited expressivity for the sake of simplification). However, sadly, {\K}'s debugger crashed on our program with the version of the {\K} framework used for this experimentation (version 3.6). This can be explained by the fact that {\K} development effort was focused on the next version to come (version 4.0 which exited the beta stage at the end of July 2016).
%
%
%
 Finally, the ability to execute the formal specification helps to validate a set of test programs that can be used as smoke test for language implementations.

\section{Conclusion}
\label{sec:conclusion}

 This paper reports on an industrial experiment to formally specify the syntax and semantics of a filtering language ({\FLang}) using the tool-supported framework {\K}. For confidentiality reasons, the filtering language specified in this report has been made up for this experimentation; however, it covers the majority of concepts usually encountered in filtering languages. No comparison between different tools is made in this experiment. The goal of the experiment is to study the feasibility of using a tool-supported formal approach for the specification of domain-specific filtering languages having a complexity similar to filtering languages encountered in real-life projects.

 The {\K} framework proved to be sufficiently expressive to naturally express the syntax and semantics of {\FLang} in a formal way. The effort required by this formal specification is judged reasonable by the authors, and within reach of average engineers which have been exposed previously to operational semantics theories. Newcomers life is made easier by the numerous manuals, examples and tutorials available for the {\K} framework. The tool support is a welcome help during the specification process. In particular, the ability to execute (or simulate) {\K} formal specifications helps greatly when developing and fine tuning the language specification, and when producing smoke tests for the implementation.

 Following such a specification process may seem to be in complete contradiction to any agile development principles~\cite{%
cockburn:2006:agile%
}.
 However, using a tool-supported \emph{executable} specification methodology allows to comply with one of the pillars of agile development: \emph{early feedback}. As the language specification is executable, it is possible to ask final users (if some are available) to test the language and provide feedbacks on different aspects of the language, including its expressivity. In fact, IBM's Continuous Engineering development methodology \cite{shamieh:2014:continuous-engineering} advocates for the use of executable models at every steps of the development.

 To summarize, with regard to the benefits of putting the effort to produce a \emph{formal} specification, the authors opinion, on improved quality and usefulness of formal specifications compared to non formal specifications written in natural language, is relatively well summarized in the following statement by David Schmidt \cite{schmidt:1996:pls}, which is supported by the numerous ambiguities (and their consequences) in natural language specifications of common programming languages like C/C++ or Java \cite{Sun:2004:JSR133}.
\begin{quote}
  ``Since data structures like symbol tables and storage vectors are explicit, a language’s subtleties are stated clearly and its flaws are exposed as awkward codings in the semantics. This helps a designer tune the language’s definition and write a better language manual. With a semantics definition in hand, a compiler writer can produce a correct implementation of the language; similarly, a user can study the semantics definition instead of writing random test programs.'' \\
  \hspace*{\stretch{1}} David Schmidt in ACM Computing Surveys \cite{schmidt:1996:pls}
\end{quote}

 In the experimentation reported in this paper, no formal analysis of the formal specification produced has been attempted. In future work, the authors plan to try some of the experimental tools available with the {\K} framework on {\FLang}'s specification. If time allows, a similar experimentation could be repeated with other tools oriented toward the formal specification of languages.

\bibliographystyle{eptcs}
\bibliography{references}

\begin{thebibliography}{10}
\providecommand{\bibitemdeclare}[2]{}
\providecommand{\surnamestart}{}
\providecommand{\surnameend}{}
\providecommand{\urlprefix}{Available at }
\providecommand{\url}[1]{\texttt{#1}}
\providecommand{\href}[2]{\texttt{#2}}
\providecommand{\urlalt}[2]{\href{#1}{#2}}
\providecommand{\doi}[1]{doi:\urlalt{http://dx.doi.org/#1}{#1}}
\providecommand{\bibinfo}[2]{#2}

\bibitemdeclare{inproceedings}{backus:1959:BNF}
\bibitem{backus:1959:BNF}
\bibinfo{author}{John~W. \surnamestart Backus\surnameend}
  (\bibinfo{year}{1959}): \emph{\bibinfo{title}{{The Syntax and Semantics of
  the Proposed International Algebraic Language of the Zurich ACM-GAMM
  Conference}}}.
\newblock In: {\sl \bibinfo{booktitle}{Proc. Int. Conf. Information
  Processing}}, \bibinfo{publisher}{UNESCO}, pp. \bibinfo{pages}{125--132}.

\bibitemdeclare{article}{basten:2015:mlir}
\bibitem{basten:2015:mlir}
\bibinfo{author}{H.~J.~S. \surnamestart Basten\surnameend},
  \bibinfo{author}{J.~\surnamestart van~den Bos\surnameend},
  \bibinfo{author}{M.~A. \surnamestart Hills\surnameend},
  \bibinfo{author}{P.~\surnamestart Klint\surnameend}, \bibinfo{author}{A.~W.
  \surnamestart Lankamp\surnameend}, \bibinfo{author}{B.~\surnamestart
  Lisser\surnameend}, \bibinfo{author}{A.~J. \surnamestart van~der
  Ploeg\surnameend}, \bibinfo{author}{T.~\surnamestart van~der
  Storm\surnameend} \& \bibinfo{author}{J.~J. \surnamestart Vinju\surnameend}
  (\bibinfo{year}{2015}): \emph{\bibinfo{title}{{Modular Language
  Implementation in Rascal -- Experience Report}}}.
\newblock {\sl \bibinfo{journal}{Science of Computer Programming}}
  \bibinfo{volume}{114}, pp. \bibinfo{pages}{7--19},
  \doi{10.1016/j.scico.2015.11.003}.

\bibitemdeclare{inproceedings}{chalub:2007:MMT}
\bibitem{chalub:2007:MMT}
\bibinfo{author}{Fabricio \surnamestart Chalub\surnameend} \&
  \bibinfo{author}{Christiano \surnamestart Braga\surnameend}
  (\bibinfo{year}{2007}): \emph{\bibinfo{title}{{Maude MSOS Tool}}}.
\newblock In: {\sl \bibinfo{booktitle}{Proc. Int. Work. Rewriting Logic and its
  Applications}}, {\sl \bibinfo{series}{Electronic Notes in Theoretical
  Computer Science}} \bibinfo{volume}{176}, \bibinfo{publisher}{Elsevier
  Science Publishers B. V.}, pp. \bibinfo{pages}{133--146},
  \doi{10.1016/j.entcs.2007.06.012}.

\bibitemdeclare{book}{cockburn:2006:agile}
\bibitem{cockburn:2006:agile}
\bibinfo{author}{Alistair \surnamestart Cockburn\surnameend}
  (\bibinfo{year}{2007}): \emph{\bibinfo{title}{{Agile Software Development:
  The Cooperative Game}}}, \bibinfo{edition}{2\textsuperscript{nd}} edition.
\newblock \bibinfo{publisher}{Pearson Education}.

\bibitemdeclare{book}{dubuisson:2000:ASN1}
\bibitem{dubuisson:2000:ASN1}
\bibinfo{author}{Olivier \surnamestart Dubuisson\surnameend}
  (\bibinfo{year}{2000}): \emph{\bibinfo{title}{{ASN.1 -- Communication between
  Heterogeneous Systems}}}.
\newblock \bibinfo{publisher}{OSS Nokalva}.
\newblock \urlprefix\url{http://www.oss.com/asn1/dubuisson.html}.
\newblock \bibinfo{note}{Translated from French by Philippe Fouquart}.

\bibitemdeclare{book}{felleisen:2009:SEwPR}
\bibitem{felleisen:2009:SEwPR}
\bibinfo{author}{Matthias \surnamestart Felleisen\surnameend},
  \bibinfo{author}{Robert~Bruce \surnamestart Findler\surnameend} \&
  \bibinfo{author}{Matthew \surnamestart Flatt\surnameend}
  (\bibinfo{year}{2009}): \emph{\bibinfo{title}{{Semantics Engineering with PLT
  Redex}}}.
\newblock \bibinfo{publisher}{The MIT Press}.

\bibitemdeclare{techreport}{flatt:2010:racket-reference}
\bibitem{flatt:2010:racket-reference}
\bibinfo{author}{Matthew \surnamestart Flatt\surnameend} \&
  \bibinfo{author}{\surnamestart {PLT}\surnameend} (\bibinfo{year}{2010}):
  \emph{\bibinfo{title}{{Reference: Racket}}}.
\newblock \bibinfo{type}{PLT-TR} \bibinfo{number}{2010-1},
  \bibinfo{institution}{PLT Design Inc.}
\newblock \bibinfo{note}{\url{https://racket-lang.org/tr1/}}.

\bibitemdeclare{inbook}{peano:1889:tpoa}
\bibitem{peano:1889:tpoa}
\bibinfo{author}{Jean \surnamestart van Heijenoort\surnameend}
  (\bibinfo{year}{2002}): \emph{\bibinfo{title}{{From Frege to Gödel: A Source
  Book in Mathematical Logic, 1879-1931}}}, chapter \bibinfo{chapter}{Peano
  (1889). The principles of arithmetic, presented by a new method}.
\newblock \bibinfo{series}{Source Books in the History of the Sciences},
  \bibinfo{publisher}{Harvard University Press}.
\newblock \bibinfo{note}{A translation and excerpt of Peano's 1889 paper
  "Arithmetices principia, nova methodo exposita"}.

\bibitemdeclare{inbook}{jezequel:2011:mdlewK}
\bibitem{jezequel:2011:mdlewK}
\bibinfo{author}{Jean-Marc \surnamestart J{\'e}z{\'e}quel\surnameend},
  \bibinfo{author}{Olivier \surnamestart Barais\surnameend} \&
  \bibinfo{author}{Franck \surnamestart Fleurey\surnameend}
  (\bibinfo{year}{2011}): \emph{\bibinfo{title}{{Summer School on Generative
  and Transformational Techniques in Software Engineering}}}, chapter
  \bibinfo{chapter}{{Model Driven Language Engineering with Kermeta}}, pp.
  \bibinfo{pages}{201--221}.
\newblock {\sl \bibinfo{series}{Lecture Notes in Computer Science}}
  \bibinfo{volume}{6491}, \bibinfo{publisher}{Springer Berlin Heidelberg},
  \doi{10.1007/978-3-642-18023-1\_5}.

\bibitemdeclare{article}{jezequel:2013:mm}
\bibitem{jezequel:2013:mm}
\bibinfo{author}{Jean-Marc \surnamestart J{\'e}z{\'e}quel\surnameend},
  \bibinfo{author}{Benoit \surnamestart Combemale\surnameend},
  \bibinfo{author}{Olivier \surnamestart Barais\surnameend},
  \bibinfo{author}{Martin \surnamestart Monperrus\surnameend} \&
  \bibinfo{author}{Fran{\c{c}}ois \surnamestart Fouquet\surnameend}
  (\bibinfo{year}{2013}): \emph{\bibinfo{title}{{Mashup of metalanguages and
  its implementation in the Kermeta language workbench}}}.
\newblock {\sl \bibinfo{journal}{Software \& Systems Modeling}}
  \bibinfo{volume}{14}(\bibinfo{number}{2}), pp. \bibinfo{pages}{905--920},
  \doi{10.1007/s10270-013-0354-4}.

\bibitemdeclare{techreport}{ISO/IEC:8824-1:2015}
\bibitem{ISO/IEC:8824-1:2015}
\bibinfo{author}{\surnamestart {Joint Technical Committee ISO/IEC JTC 1,
  Information technology, Subcommittee SC 6, Telecommunications and information
  exchange between systems}\surnameend} (\bibinfo{year}{2015}):
  \emph{\bibinfo{title}{{Information technology -- Abstract Syntax Notation One
  (ASN.1): Specification of basic notation}}}.
\newblock \bibinfo{type}{International Standard} \bibinfo{number}{8824-1},
  \bibinfo{institution}{ISO/IEC}.
\newblock \bibinfo{note}{{ISO/IEC} version of ITU-T X.680 (08/2015)}.

\bibitemdeclare{techreport}{Sun:2004:JSR133}
\bibitem{Sun:2004:JSR133}
\bibinfo{author}{\surnamestart {JSR-133 expert group}\surnameend}
  (\bibinfo{year}{2004}): \emph{\bibinfo{title}{{JSR-133 Java{\texttrademark}
  Memory Model and Thread Specification Revision}}}.
\newblock \bibinfo{type}{Java Specification Request (JSR)}
  \bibinfo{number}{133}, \bibinfo{institution}{{Sun Microsystems, Inc.}}
\newblock \bibinfo{note}{\url{https://jcp.org/en/jsr/detail?id=133}}.

\bibitemdeclare{inproceedings}{klein:2012:ryr}
\bibitem{klein:2012:ryr}
\bibinfo{author}{Casey \surnamestart Klein\surnameend}, \bibinfo{author}{John
  \surnamestart Clements\surnameend}, \bibinfo{author}{Christos \surnamestart
  Dimoulas\surnameend}, \bibinfo{author}{Carl \surnamestart
  Eastlund\surnameend}, \bibinfo{author}{Matthias \surnamestart
  Felleisen\surnameend}, \bibinfo{author}{Matthew \surnamestart
  Flatt\surnameend}, \bibinfo{author}{Jay~A. \surnamestart
  McCarthy\surnameend}, \bibinfo{author}{Jon \surnamestart Rafkind\surnameend},
  \bibinfo{author}{Sam \surnamestart Tobin-Hochstadt\surnameend} \&
  \bibinfo{author}{Robert~Bruce \surnamestart Findler\surnameend}
  (\bibinfo{year}{2012}): \emph{\bibinfo{title}{{Run Your Research: On the
  Effectiveness of Lightweight Mechanization}}}.
\newblock In: {\sl \bibinfo{booktitle}{Proc. Symp. Principles of Programming
  Languages}}, {\sl \bibinfo{series}{SIGPLAN Not.}}~\bibinfo{volume}{47},
  \bibinfo{publisher}{ACM}, \bibinfo{address}{New York, NY, USA}, pp.
  \bibinfo{pages}{285--296}, \doi{10.1145/2103656.2103691}.

\bibitemdeclare{incollection}{klint:2009:t2gmt}
\bibitem{klint:2009:t2gmt}
\bibinfo{author}{P.~\surnamestart Klint\surnameend} (\bibinfo{year}{2009}):
  \emph{\bibinfo{title}{{Tribute to a great Meta-Technologist: from Centaur to
  The Meta-Environment}}}.
\newblock In \bibinfo{editor}{Y.~\surnamestart Bertot\surnameend},
  \bibinfo{editor}{G.~\surnamestart Huet\surnameend}, \bibinfo{editor}{J.-J.
  \surnamestart Levy\surnameend} \& \bibinfo{editor}{G.~\surnamestart
  Plotkin\surnameend}, editors: {\sl \bibinfo{booktitle}{From Semantics to
  Computer Science, Essays in Honour of Gilles Kahn}},
  \bibinfo{publisher}{Cambridge University Press}, pp.
  \bibinfo{pages}{235--264}, \doi{10.1017/CBO9780511770524.012}.

\bibitemdeclare{inproceedings}{klint:2011:RLSRunner}
\bibitem{klint:2011:RLSRunner}
\bibinfo{author}{P.~\surnamestart Klint\surnameend}, \bibinfo{author}{J.~J.
  \surnamestart Vinju\surnameend} \& \bibinfo{author}{M.~A. \surnamestart
  Hills\surnameend} (\bibinfo{year}{2011}): \emph{\bibinfo{title}{{RLSRunner:
  Linking Rascal with K for Program Analysis}}}.
\newblock In: {\sl \bibinfo{booktitle}{Proc. Int. Conf. Software Language
  Engineering}}, \bibinfo{publisher}{Springer},
  \doi{10.1007/978-3-642-28830-2\_19}.

\bibitemdeclare{inproceedings}{klint:2009:RASCAL}
\bibitem{klint:2009:RASCAL}
\bibinfo{author}{Paul \surnamestart Klint\surnameend}, \bibinfo{author}{Tijs
  \surnamestart van~der Storm\surnameend} \& \bibinfo{author}{Jurgen
  \surnamestart Vinju\surnameend} (\bibinfo{year}{2009}):
  \emph{\bibinfo{title}{{RASCAL: A Domain Specific Language for Source Code
  Analysis and Manipulation}}}.
\newblock In: {\sl \bibinfo{booktitle}{Proc. Int. Working Conf. Source Code
  Analysis and Manipulation}}, \bibinfo{publisher}{IEEE Computer Society}, pp.
  \bibinfo{pages}{168--177}, \doi{10.1109/SCAM.2009.28}.

\bibitemdeclare{article}{knuth:1964:BNFvsBNF}
\bibitem{knuth:1964:BNFvsBNF}
\bibinfo{author}{Donald~E. \surnamestart Knuth\surnameend}
  (\bibinfo{year}{1964}): \emph{\bibinfo{title}{{Backus Normal Form vs. Backus
  Naur Form}}}.
\newblock {\sl \bibinfo{journal}{Commun. ACM}}
  \bibinfo{volume}{7}(\bibinfo{number}{12}), pp. \bibinfo{pages}{735--736},
  \doi{10.1145/355588.365140}.

\bibitemdeclare{techreport}{le-guernic:2016:erfspflukf}
\bibitem{le-guernic:2016:erfspflukf}
\bibinfo{author}{Gurvan \surnamestart Le~Guernic\surnameend} \&
  \bibinfo{author}{José~A. \surnamestart Galindo\surnameend}
  (\bibinfo{year}{2016}): \emph{\bibinfo{title}{{Experience Report on the
  Formal Specification of a Packet Filtering Language Using the K Framework}}}.
\newblock \bibinfo{type}{Research report} \bibinfo{number}{8967},
  \bibinfo{institution}{Inria}.
\newblock \bibinfo{note}{\url{https://hal.inria.fr/hal-01385541v1}}.

\bibitemdeclare{inproceedings}{mulligan:2014:Lem}
\bibitem{mulligan:2014:Lem}
\bibinfo{author}{Dominic~P. \surnamestart Mulligan\surnameend},
  \bibinfo{author}{Scott \surnamestart Owens\surnameend},
  \bibinfo{author}{Kathryn~E. \surnamestart Gray\surnameend},
  \bibinfo{author}{Tom \surnamestart Ridge\surnameend} \&
  \bibinfo{author}{Peter \surnamestart Sewell\surnameend}
  (\bibinfo{year}{2014}): \emph{\bibinfo{title}{{Lem: Reusable Engineering of
  Real-world Semantics}}}.
\newblock In: {\sl \bibinfo{booktitle}{Proc. Int. Conf. Functional
  Programming}}, {\sl \bibinfo{series}{SIGPLAN Not.}}~\bibinfo{volume}{49},
  \bibinfo{publisher}{ACM}, pp. \bibinfo{pages}{175--188},
  \doi{10.1145/2692915.2628143}.

\bibitemdeclare{article}{rosu:2010:K-overview}
\bibitem{rosu:2010:K-overview}
\bibinfo{author}{Grigore \surnamestart Ro{\c s}u\surnameend} \&
  \bibinfo{author}{Traian~Florin \surnamestart {\c S}erb{\u a}nu{\c t}{\u
  a}\surnameend} (\bibinfo{year}{2010}): \emph{\bibinfo{title}{{An Overview of
  the $\mathbb{K}$ Semantic Framework}}}.
\newblock {\sl \bibinfo{journal}{The Journal of Logic and Algebraic
  Programming}} \bibinfo{volume}{79}(\bibinfo{number}{6}), pp.
  \bibinfo{pages}{397--434}, \doi{10.1016/j.jlap.2010.03.012}.

\bibitemdeclare{inproceedings}{rosu:2014:K-overview}
\bibitem{rosu:2014:K-overview}
\bibinfo{author}{Grigore \surnamestart Roşu\surnameend} \&
  \bibinfo{author}{Traian~Florin \surnamestart Şerbănuţă\surnameend}
  (\bibinfo{year}{2014}): \emph{\bibinfo{title}{{$\mathbb{K}$ Overview and
  SIMPLE Case Study}}}.
\newblock In: {\sl \bibinfo{booktitle}{Proc. Int. Work. K Framework and its
  Applications (K 2011)}}, {\sl \bibinfo{series}{Electronic Notes in
  Theoretical Computer Science}} \bibinfo{volume}{304}, pp.
  \bibinfo{pages}{3--56}, \doi{10.1016/j.entcs.2014.05.002}.

\bibitemdeclare{article}{schmidt:1996:pls}
\bibitem{schmidt:1996:pls}
\bibinfo{author}{David~A. \surnamestart Schmidt\surnameend}
  (\bibinfo{year}{1996}): \emph{\bibinfo{title}{{Programming Language
  Semantics}}}.
\newblock {\sl \bibinfo{journal}{ACM Computing Surveys}}
  \bibinfo{volume}{28}(\bibinfo{number}{1}), \doi{10.1145/234313.234419}.

\bibitemdeclare{article}{sewell:2010:Ott}
\bibitem{sewell:2010:Ott}
\bibinfo{author}{Peter \surnamestart Sewell\surnameend},
  \bibinfo{author}{Francesco~Zappa \surnamestart Nardelli\surnameend},
  \bibinfo{author}{Scott \surnamestart Owens\surnameend},
  \bibinfo{author}{Gilles \surnamestart Peskine\surnameend},
  \bibinfo{author}{Thomas \surnamestart Ridge\surnameend},
  \bibinfo{author}{Susmit \surnamestart Sarkar\surnameend} \&
  \bibinfo{author}{Rok \surnamestart Strni\v{s}a\surnameend}
  (\bibinfo{year}{2010}): \emph{\bibinfo{title}{{Ott: Effective Tool Support
  for the Working Semanticist}}}.
\newblock {\sl \bibinfo{journal}{J. Functional Programming}}
  \bibinfo{volume}{20}(\bibinfo{number}{1}), pp. \bibinfo{pages}{71--122},
  \doi{10.1017/S0956796809990293}.

\bibitemdeclare{book}{shamieh:2014:continuous-engineering}
\bibitem{shamieh:2014:continuous-engineering}
\bibinfo{author}{Cathleen \surnamestart Shamieh\surnameend}
  (\bibinfo{year}{2014}): \emph{\bibinfo{title}{{Continuous Engineering For
  Dummies\textsuperscript{\textregistered}}}}.
\newblock \bibinfo{series}{IBM Limited Edition}, \bibinfo{publisher}{John Wiley
  \& Sons, Inc}.

\bibitemdeclare{techreport}{storm:2008:Meta-Env}
\bibitem{storm:2008:Meta-Env}
\bibinfo{author}{T.~\surnamestart van~der Storm\surnameend} \&
  \bibinfo{author}{J.~J. \surnamestart Vinju\surnameend}
  (\bibinfo{year}{2008}): \emph{\bibinfo{title}{{Using the Meta-Environment for
  Domain Specific Language Engineering}}}.
\newblock \bibinfo{type}{Technical Report} \bibinfo{number}{SEN-R0805},
  \bibinfo{institution}{CWI Software Engineering}.

\bibitemdeclare{article}{serbanuta:2014:K-primer}
\bibitem{serbanuta:2014:K-primer}
\bibinfo{author}{Traian~Florin \surnamestart Şerbănuţă\surnameend},
  \bibinfo{author}{Andrei \surnamestart Arusoaie\surnameend},
  \bibinfo{author}{David \surnamestart Lazar\surnameend},
  \bibinfo{author}{Chucky \surnamestart Ellison\surnameend},
  \bibinfo{author}{Dorel \surnamestart Lucanu\surnameend} \&
  \bibinfo{author}{Grigore \surnamestart Roşu\surnameend}
  (\bibinfo{year}{2014}): \emph{\bibinfo{title}{{The $\mathbb{K}$ Primer
  (version 3.3)}}}.
\newblock {\sl \bibinfo{journal}{Electronic Notes in Theoretical Computer
  Science}} \bibinfo{volume}{304}, pp. \bibinfo{pages}{57--80},
  \doi{10.1016/j.entcs.2014.05.003}.
\newblock \bibinfo{note}{Proc. Int. Work. K Framework and its Applications (K
  2011)}.

\end{thebibliography}

\end{document}